\documentclass[a4paper,fleqn,usenatbib]{mnras}

\usepackage[T1]{fontenc}
\usepackage{ae,aecompl}

\usepackage{graphicx}	
\usepackage{amsmath}	
\usepackage{amssymb}	
\usepackage{subfig}
\usepackage{hyperref}
\usepackage{soul}

\title[Broadband study of PKS 0208-512]{Temporal and spectral study of PKS\,0208-512 during the 2019-2020 flare}

\author[R. Khatoon et al.]{
Rukaiya Khatoon$^{1,2}$\thanks{E-mail: rukaiyakhatoon12@gmail.com},
Raj Prince$^{3}$, Zahir Shah$^{2,4}$, Sunder Sahayanathan$^{5,6}$ \newauthor and Rupjyoti Gogoi$^{1}$  
\\ \\
$^{1}$Tezpur University,Napaam-784028, Assam, India.\\
$^{2}$Inter-University Center for Astronomy and Astrophysics, Post Bag 4, Ganeshkhind, Pune-411007, India. \\
$^{3}$Center for Theoretical Physics, Polish Academy of Sciences, Al.Lotnikov 32/46, PL-02-668 Warsaw, Poland.\\
$^{4}$Department of Physics, Central University of Kashmir, India.\\
$^{5}$Astrophysical Sciences Division, Bhabha Atomic Research Centre, Mumbai-400085, India.\\
$^{6}$Homi Bhabha National Institute, Mumbai 400094, India.
}

\date{Accepted XXX. Received YYY; in original form ZZZ}

\pubyear{2020}

\begin{document}
\label{firstpage}
\pagerange{\pageref{firstpage}--\pageref{lastpage}}
\maketitle


\begin{abstract}

We present the temporal and spectral study of blazar PKS\,0208-512, using recent flaring activity from November 2019 to May 2020, as detected by the {\it Fermi}-LAT observatory. The contemporaneous X-ray, optical/UV observations from \emph{Swift}-XRT/UVOT are also used. During the activity state, 2-days binned $\gamma$-ray lightcurve shows multiple peaks indicating sub-flares. To understand the possible physical mechanisms behind flux enhancement, we divided the activity state of the source into several flux-states and carried out detailed temporal and spectral studies. Timing analysis of lightcurves suggest that peaks of sub-flares have rise and decay time on the order of days, with flux-doubling time $\sim$ 2-days. The 2-days binned $\gamma$-ray lightcurve shows double-lognormal flux distribution. The broadband spectral-energy-distribution (SED) for three selected flux states can be well fitted under synchrotron, synchrotron self-Compton (SSC) and external-Compton (EC) emission mechanisms. We obtained the physical parameters of the jet by the SED modeling and their confidence intervals through $\chi^{2}$-statistics. Our SED modeling results suggest that during quiescent-state, $\gamma$-ray spectrum can be well explained by considering the EC-scattering of infra-red (IR) photons from dusty-torus. However, $\gamma$-ray spectra corresponding to flares demand additional target photons from broad-line-region (BLR) along with IR. These suggest that during flares, the emission-region is close to the edge of BLR, while for quiescent-state the emission-region is away from BLR. The best fit results suggest that marginal increase in magnetic-field during the flaring-episode can result in the flux enhancement. This is possibly associated with the efficiency of particle acceleration during flaring-states as compared to quiescent-state.

\end{abstract}



\section{Introduction}
Blazars are the class of active galactic nuclei (AGN) with relativistic jets pointing towards the observer within a few degrees (\citealt{Urry_1995}).
The jet emission is predominantly non-thermal in nature covering
the entire electromagnetic (EM) spectrum ranging from radio to very high energy $\gamma$-ray. 
Their broadband spectral energy distribution (SED) is characterized by two prominent peaks, with the low energy peak generally observed between infrared (IR) to soft X-ray, while the high energy one falls at Mev/GeV energies. The low energy component is well explained by the synchrotron emission from relativistic electrons spiraling along the magnetic fields in the jet. 
The high energy component is generally modelled as the inverse-Compton (IC) scattering of low energy photons by the relativistic electrons. The source of low energy 
target photons is either internal (synchrotron emission) or external (accretion disk, broad-line region,  dusty torus, etc.) to the jet (\citealt{Bottcher_97, Ghisellini_96, Blazejowsi_2000}). Such a model is commonly referred as a leptonic model (\citealt{2009MNRAS.397..985G, 2014Natur.515..376G}). If the target photons for the inverse-Compton scattering is synchrotron photons itself
then the process is referred to as synchrotron self-Compton (SSC; \citealt{Sikora_2009}). On the other hand, the scattering of external photons is known as external-Compton (EC; \citealt{Dermer_92}; \citealt{Sikora_94}; \citealt{10.1046/j.1365-8711.1998.02032.x}) process. \\

One of the intriguing properties of blazars is that they show strong stochastic variability and occasional strong flares across the EM spectrum. The flux doubling time scale varies from minutes to several days  (\citealt{Paliya_2015}, \citealt{2020NatCo..11.4176S}, \citealt{2021JApA...42...80C}). The rapid flux variability of the order of minutes/hours suggests the emission is produced from a region very close to the central engine (\citealt{2008MNRAS.386L..28G, narayan}). 
In the past three decades, development in ${\gamma}$-ray astronomy has been an important tool to understand the spectral and temporal behavior of blazars (\citealt{1995ApJ...440..525V, Abdo, Abdo_2010, 2016ARA&A..54..725M}). Particularly, the launch of the {\it Fermi}-Large Area Telescope (LAT, \citealt{Atwood_2009}) opened up a new window to study high energy $\gamma$-ray blazars. The recently updated {\it Fermi}-LAT catalog, 4FGL (\citealt{2020ApJS..247...33A}) has more than five thousand $\gamma$-ray sources, 95\% of which are blazars.  A detailed study of the $\gamma$-ray blazars will help us to understand the acceleration and the dynamics of the underlying particle distribution. However, this also demands additional simultaneous broadband
observation of the source in tandem with $\gamma$-rays.
In the recent past, various telescopes i.e., the Neil Gehrels \emph{Swift} observatory (\citealt{2004ApJ...611.1005G}), \emph{Steward} observatory (\citealt{2009arXiv0912.3621S}), Owens Valley Radio Observatory (\emph{OVRO}) 40-m (\citealt{Richards_2011}) etc. operating at different energy ranges have been used to observe blazars along with 
{\it Fermi}-LAT. The simultaneous or quasi-simultaneous observations available from these instruments were collectively used to study the complex nature of 
blazars (\citealt{Prince_2019,2021MNRAS.502.5245P,2021MNRAS.tmp..823S}). 

The flat spectrum radio quasar (FSRQ) PKS\,0208-512 is located at redshift z = 1.003 (\citealt{2008ApJS..175...97H}). The Parkes radio survey discovered the source (\citealt{1964AuJPh..17..340B}) and was observed regularly in ${\gamma}$-rays (\citealt{1995ApJS..101..259T}) by \emph{EGRET} onboard Compton Gamma-Ray Observatory (\emph{CGRO}). The \emph{CGRO} mission detected excess emission from PKS\,0208-512 at soft $\gamma$-rays (1-3 MeV) during 1993 May-June. The $\gamma$-ray SED indicated a broad peak at MeV energies (\citealt{1995A&A...298L..33B}), later confirmed by INTEGRAL/SPI detection (\citealt{2010A&A...514A..69Z}). Henceforth, the source has been identified as `MeV blazar'.

PKS 0208-512 has been monitored regularly by the {\it Fermi}-LAT in a wide range 0.1-300 GeV and during October 2008 a moderate $\gamma$-ray flare was detected (\citealt{2008ATel.1759....1C}). The contemporaneous observations in optical/IR during this $\gamma$-ray flare, has been observed by \emph{SMARTS} and \emph{ANDICAM} (\citealt{2008ATel.1751....1B}). Subsequently, several studies on the multiwavelength property of PKS\,0208-512 have also been undertaken (\citealt{2010MNRAS.402..497G, 2010A&A...514A..69Z,2011ATel.3338....1S, 2011ATel.3421....1N,2012arXiv1202.4499B,2013ApJ...763L..11C,2013ApJ...771L..25C}).

From November, 2019 to May, 2020, PKS\,0208-512 went through a major $\gamma$-ray flare 
and the follow-up ATel initiated a major coordinated observation of the 
source (\citealt{2019ATel13320....1A,2019ATel13352....1L,2020ATel13558....1A,2020ATel13847....1A}). 
The preliminary analysis of {\it Fermi}-LAT observations suggested that the daily averaged $\gamma$-ray flux 
at energy >100MeV
enhanced up to 1.1$\pm$0.2 $\times10^{-6}$ ph\,cm$^ {-2}$\,s$^{-1}$ on November 29, 2019, and 
2.0$\pm$0.3 $\times10^{-6}$ ph\,cm$^{-2}$\,s$^{-1}$ on March 15, 2020 \citep{2019ATel13320....1A,2020ATel13558....1A}. Further, \emph{AGILE} observed the source during December 14-16, 2019 and using multi-source maximum likelihood analysis it measured a flux of 
2.7$\pm$0.8 $\times10^{-6}$ ph\,cm$^{-2}$\,s$^{-1}$ for energy >100 MeV (\citealt{2019ATel13352....1L}).

The follow-up observations in the optical/UV and X-ray 
bands have been performed by \emph{Swift} telescope under the Target of opportunity (ToO) program during December 17-24, 2019,  
and March 9-21, 2020. The rich simultaneous multi-wavelength observation of PKS\,0208-512 during different flaring activity  
encouraged us to perform a detailed broadband timing and spectral study of the source. Our primary study is based on the 
$\gamma$-ray observation by {\it Fermi}-LAT and we supplement this with the simultaneous observation by \emph{Swift}-XRT/UVOT
in X-ray and optical/UV wavebands.
The paper is organised as follows: In \S 2, we describe the multi-wavelength observations and data analysis procedure.
In \S 3.1, we provide the broadband lightcurve and the identification of different flux states followed by the results 
obtained from the temporal and spectral analysis of $\gamma$-ray observations (\S 3.2 -- \S 3.5). 
The results from \emph{Swift} observations 
are given in \S 3.6. The detailed broadband spectral modeling of the source using synchrotron and inverse Compton emission
processes is described in \S 4. Finally, the study results are discussed and summarized in \S 5. 
Throughout this work, we adopt a cosmology with $\Omega_m = 0.3$, $\Omega_\Lambda = 0.7$
and $H_0 = 70$ km s$^{-1}$ Mpc$^{-1}$.

\section{Observations and data analysis}

\subsection{{\it Fermi}-LAT}
LAT (\citealt{Atwood_2009}) is a $\gamma$-ray instrument onboard {\it Fermi} satellite launched by NASA in 2008 and sensitive in the energy range between 20 MeV to >300 GeV. It has a large field of view $\sim$ 2.4 sr and takes approximately 3 hours to scan the entire sky. The source PKS\,0208-512, has been regularly  monitored by {\it Fermi}-LAT. To study the long term $\gamma$-ray behavior of the source, we analyzed the data from September 4, 2017 to July 20,
2020 (MJD 58000-59050). This period includes both the low flux states as well as the high flux states of the source. The analysis has been carried out with the 10 degree region of interest (ROI) centered at the source position. We followed the standard analysis procedure described by the Fermi science tools \footnote{https://fermi.gsfc.nasa.gov/ssc/data/analysis/documentation/} (version v10r0p5). The analysis was performed in the energy range from 100 MeV to 300 GeV with the evclass=128 and evtype=3, and a zenith angle cut >90$^{\circ}$ is applied to reduce the contamination from the Earth's limb $\gamma$-rays. To generate the model xml file, we used the galactic diffuse emission model, ``\emph{$gll_-iem_-v07$}" and the isotropic background model, ``\emph{$iso_-P8R3_-SOURCE_-V2$}". The model files are available publicly at the Fermi Science Support Center (FSSC). The spectral models and the spectral parameters for the sources in the ROI are defined in the fourth Fermi source catalog (4FGL; \citealt{2020ApJS..247...33A}). To optimize the spectral parameters of the sources in the ROI, maximum Likelihood method is used and the detection of these sources were quantified by the test statistics (TS) \footnote{To quantify the significance of the source, we used the Likelihood ratio test statistic (TS) defined as TS = 2 log(L$_1$/L$_0$), where L$_1$ and L$_0$ are the maximum Likelihood values for a model with and without the source in the specified location. The significance of the source 
detection is determined by ${\sim}$ TS$^{(1/2)}$ ${\times}{\sigma}$}. The sources with TS < 9 (corresponds to $\sim$ 3${\sigma}$; \citealt{Mattox_1996}) have been removed for further analysis. The above procedure has been followed using the ``unbinned likelihood analysis with python" developed by the Fermi collaboration. 

The default spectral model of the source PKS 0208-512 is log parabola, the model parameters are optimized by the likelihood analysis.  Finally, we fit the source spectrum integrated over the entire energy range 0.1-300 GeV during period (MJD 58000-59050) with a log-parabola function defined as
\begin{equation}
    N(E)\,dE = N_0\times\left(\frac{E}{E_0}\right)^{-(\alpha+\beta \, \text{log}(E/E_0))} dE
\end{equation}

where, $\alpha$ is the spectral slope at reference energy $E_0$ and $\beta$ is the peak spectral curvature. 
During the lightcurve generation, we freeze the parameters of all the sources except our source of interest PKS\,0208-512, and 2-days binned $\gamma$-ray lightcurves for the flux and spectral parameters have been produced.

\subsection{\emph{Swift}-XRT}
\emph{Swift} is a space-based telescope also known as Neil Gehrels observatory, was launched in November 2004 to 
observe the transient phenomenon in the Universe (\citealt{2004ApJ...611.1005G}). The instruments onboard \emph{Swift}, used in our study are  X-ray telescope (XRT, 0.3-10 keV), ultraviolet optical telescope (UVOT, 170-650 nm) with filters V, B, U for optical band and W1, M2, and W2 for UV band, respectively (\citealt{Larionov_2016}). 
The source PKS\,0208-512, has been observed regularly through the monitoring program and also by the target of opportunity (ToO) 
program. During the period MJD 58000-59050, Swift observed the source PKS 0208-512 in 
X-ray, optical, and UV as part of a ToO program only. The Swift ToO observations are follow-up of 
the $\gamma$-ray flaring activity of the source, reported on December 1, 2019 and March 16, 2020 (\citealt{2019ATel13320....1A,2020ATel13558....1A}). The follow-up Swift ToO observations were performed during December 17-24, 2019 (observation IDs : 00041512021, 00041512023, 00041512025, 00041512026, and 00041512027) and March 9-21, 2020 (observation IDs : 00035002131, 00035002132, 00035002133, 00035002134, and 00035002135). In total, we have 12 \emph{Swift} observations of $\sim$ 1-2 ks each, for a total exposure time of 18.8 ks during the period from MJD 58669 to 58930. To reduce the X-ray data, we have followed the standard procedure. First, we ran the \texttt{XRTPIPELINE} to produce the clean event files, and to do that, the latest version of the calibration file (CALDB, version 20200305) is used. The cleaned event files for the photon counting (PC) mode were further used to create a source and background spectrum, using \texttt{XSELECT}. 

To extract the source spectrum, a circular region of 20 arcsec is chosen around the source, while for the background spectrum, a circular region of 40 arcsec is selected far from the source to avoid contamination. In the case of pile-up affected observations with the source count rate above 0.5 ct/sec, an annular region of 4 arcsec inner radius and 20 arcsec outer radius were considered as the source region.  The ancillary response files (ARF) and the redistribution matrix files (RMF) were generated using \texttt{XRTMKARF} from the \emph{Swift} CALDB.  For the flaring state, there are multiple observations in a given period and spectra from all these observations were combined using addspec\footnote{https://heasarc.gsfc.nasa.gov/ftools/caldb/help/addspec.txt}. In addspec, RMF and  ARF files for different observations were added with the source spectrum. The background spectra from different observations were added through mathpha\footnote{https://heasarc.gsfc.nasa.gov/ftools/caldb/help/mathpha.txt}. Finally, the source spectrum and the background spectrum were combined through the tool \texttt{grppha}, and 20 counts in each bin have been considered. The resultant grouped spectrum was then added to the \texttt{XSPEC} for the modeling. We fit the X-ray spectrum corresponding to each ToO observation with an absorbed power-law model using \texttt{XSPEC} (\citealt{1996ASPC..101...17A}), and estimate the flux and index. The Swift-XRT lightcurve is generated with each point corresponding to 
one observation. The X-ray spectrum obtained during the high and low state was used for the broadband SED modeling.

\subsection{\emph{Swift}-UVOT}
\emph{Swift}-UVOT (\citealt{Roming_2005}) covers the optical and UV part of the spectrum with its three optical (U, B, V) and three UV (W1, M2, W2) filters. In case of multiple observations for a given period, we combined the images in the filters using the \texttt{UVOTIMSUM} tool. To extract the magnitudes from the images, the task \texttt{UVOTSOURCE} has been used with the source and background regions of 5 arcsec and 10 arcsec, respectively. The observed magnitudes were then corrected for the Galactic extinction, with E(B - V) = 0.0174 mag and the ratio A$_{V}$ /E(B - V ) = 3.1 from \citet{Schlafly_2011}. We converted the magnitudes to flux units using the photometric zero-points from \citet{Breeveld_2011} and the conversion factors from \citet{Larionov_2016}.

\section{Results}

{\it Fermi}-LAT observation of PKS\,0208-512 has shown enhanced flaring activity in $\gamma$-ray during the period 
2019-2020 \citep{2019ATel13320....1A,2020ATel13558....1A}.
A daily averaged $\gamma$-ray flux (E>100MeV) of (1.1$\pm$0.2)$\times10^{-6}$ ph\,cm$^{-2}$\,s$^{-1}$ was reported for this source on November 29, 2019
(\citealt{2019ATel13320....1A}). Subsequently, an increased flux with 2.0$\pm$0.3 $\times10^{-6}$ ph\,cm$^{-2}$\,s$^{-1}$ was observed on March 15, 2020, 
which is the highest daily binned $\gamma$-ray flux ever detected from this source (\citealt{2020ATel13558....1A}).
In the 2-days binned $\gamma$-ray lightcurve extending from September 4, 2017 to July 20,
2020 (MJD 58000-59050) and integrated over the energy range 100 MeV - 300 GeV, the peak 
flux (1.65$\pm$0.15) $\times10^{-6}$ ph\,cm$^{-2}$\,s$^{-1}$ is observed on March 15, 2020. This flux value is slightly lower than the daily binned flux reported by \citealt{2020ATel13558....1A}. Moreover, we noted that 2-days binned $\gamma$-ray flux detected by {\it Fermi}-LAT is $\sim$ (1.43$\pm$0.14)$\times10^{-6}$ ph\,cm$^{-2}$\,s$^{-1}$ on December 16, 2019, which is lower than the 2-days integrated flux $\sim$ (2.7$\pm$0.8)$\times10^{-6}$ ph\,cm$^{-2}$\,s$^{-1}$ detected by AGILE on the same day. Such discrepancy in flux value can be possibly associated with the detection from different instruments and the analysis procedure.
The maximum fluxes in X-ray, optical (B-band) and UV  (W1-band), overlapping with the $\gamma$-ray lightcurve, are found to be 
(1.8$\pm$0.2) $\times10^{-11}$ erg cm$^ {-2}$ s$^{-1}$, (2.14$\pm$0.06) $\times10^{-11}$ erg cm$^ {-2}$ s$^{-1}$ 
and (1.65$\pm$0.05) $\times10^{-11}$ erg cm$^ {-2}$ s$^{-1}$ falling on March 9, 2020 as indicated by Swift-XRT/UVOT analysis. 
Here, we perform a detailed temporal and spectral study of the brightest $\gamma$-ray flare witnessed during MJD 58780-59000.


\subsection{Multiwavelength lightcurves}

\begin{figure*}
        \centering
        \includegraphics[scale=0.6]{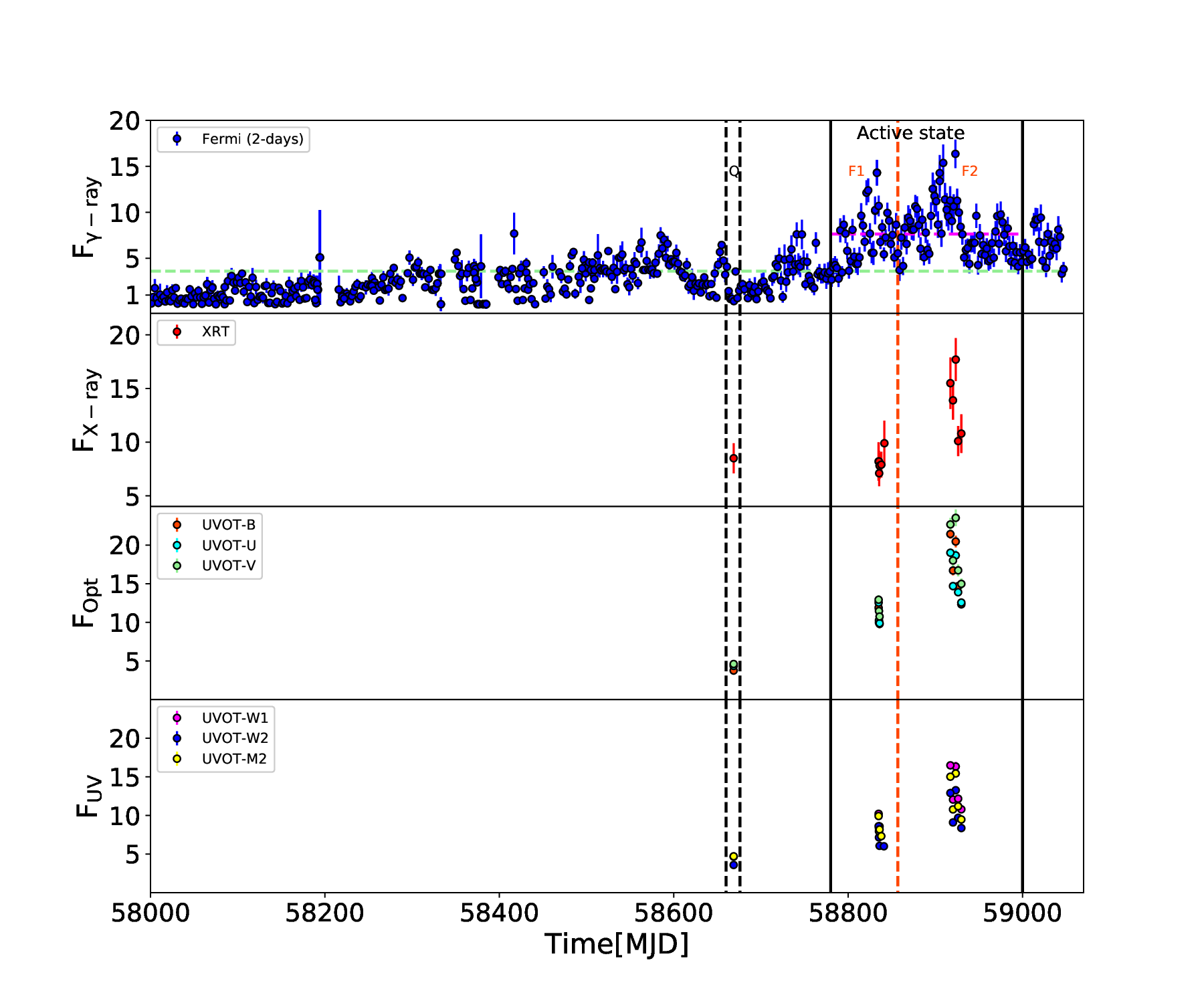}
        \caption{Multiwavelength lightcurves of PKS\,0208-512 between MJD 58000 - 59050. Panel 1: {\it Fermi}-LAT flux in $10^{-7} ph/cm^2/sec$ averaged over 2-days. Panel 2: \emph{Swift}-XRT flux in $10^{-12} erg/cm^2/sec$. Panel 3 \& 4: Optical and UV flux in $10^{-12} erg/cm^2/sec$ (in the \emph{Swift}-UVOT B, V, U bands, and \emph{Swift}-UVOT W1, W2, M2 bands). The vertical dashed lines in black color denote the Quiescent state (Q-state), and Active state of PKS\,0208-512. The vertical dashed line in red color is used to represents two flaring states (F1 $\&$ F2 states). The horizontal dashed line in green color represents the average of the 2-days binned $\gamma$-ray flux, and the horizontal dashed line in magenta color represents the average of the flux in the Active state.}
		
        \label{fig:mwl_lc}
\end{figure*}

The $\gamma$-ray lightcurve during the period from September 4, 2017 to July 20, 2020 (MJD 58000-59050), is shown in the top panel of Figure \ref{fig:mwl_lc}, 
where green dashed horizontal line corresponds to the average flux F$_{b}$ = 3.60$\times10^{-7}$ ph\,cm$^ {-2}$\,s$^{-1}$, obtained from the 2-days binned data. 
The lightcurve in comparison with F$_b$ suggests that the source was in a low flux state for a long period during September 4, 2017 to October 24, 2019 (MJD 58000-58780) with 
occasional minor outbursts. The state of enhanced $\gamma$-ray emission was observed from October 24, 2019 to May 31, 2020 (MJD 58780 - 59000), with 
the fluxes rising significantly above F$_{b}$. We consider this period as the `active state' of the source and demarcate it with vertical black bold
lines in Figure \ref{fig:mwl_lc}. 
During the $\gamma$-ray active state, two giant flares were identified with the first one peaking on December 16, 2019 (MJD 58833) with flux (1.43$\pm$0.14)$\times10^{-6}$ ph\,cm$^ {-2}$\,s$^{-1}$ 
and the second one peaking on March 15, 2020 (MJD 58923) with flux (1.65$\pm$0.15) $\times10^{-6}$ ph\,cm$^ {-2}$\,s$^{-1}$. The average flux during the active state is 
F$_{a}$ = 7.6 $\times10^{-7}$ ph\,cm$^ {-2}$\,s$^{-1}$ which is shown as dashed horizontal magenta line. Based on F$_a$, we divide the active region into two flaring 
states namely flare-1 (F1) and flare-2 (F2) with fluxes significantly higher than F$_{a}$.
The peaks of these two flares are separated by $\sim$ 90 days. We have demarcated this division by red dashed 
vertical line in Figure \ref{fig:mwl_lc}. 
The second, third, and fourth panels from the top of Figure \ref{fig:mwl_lc} correspond to X-ray, optical, and UV observations. It can be noted that the high flux 
points in X-ray, optical, and UV bands fall within the active state of $\gamma$-ray. Similarly, during $\gamma$-ray low flux state on July 4, 2019 (MJD 58668) we have single observation in other energy bands. The fluxes during this period are (6.64$\pm$2.64)$\times10^{-8}$ ph\,cm$^ {-2}$\,s$^{-1}$ in $\gamma$-ray,  (8.5$\pm$1.2)$\times10^{-12}$ erg cm$^ {-2}$ s$^{-1}$ in X-ray, (3.76$\pm$0.27)$\times10^{-12}$ erg cm$^ {-2}$ s$^{-1}$ in B-band, and (4.67$\pm$0.3)$\times10^{-12}$ erg cm$^ {-2}$ s$^{-1}$ in W1-band, respectively. Consistently, the flux values in X-ray, optical, and UV bands are minimal compared to ones observed during 
the $\gamma$-ray active state. We denote the $\gamma$-ray low flux state with simultaneous to X-ray, optical, and UV observations during the period from June to July, 2019 (MJD 58660 - 58676) as quiescent state `Q' and represented by black dashed vertical lines.

\subsection{$\gamma$-ray variability study}
The variability behavior of the source can be quantified through fractional RMS variability amplitude, F$_{var}$(\citealt{Edelson_2002}) which
is defined as the squared root of excess  variance  ($\sigma_{XS}$). It accounts for the variability induced by the measurement uncertainties and its functional form can be represented as \citep{Vaughan_2003}
\begin{equation}\label{eq:fvar}
 F_{\rm var} = \sqrt{\frac{S^2 - err^2}{F^2}},
\end{equation}
Here, {\it F} and {\it err$^2$} are the average flux and the mean square error in the observed flux, and the 
total variance of the observed lightcurve is denoted by {\it  S$^2$}. The error on {\it F$_{var}$} can be estimated as \citep{Vaughan_2003}
\begin{equation}
err(F_{var}) = \sqrt{ \Big(\sqrt{\frac{1}{2N}}. \frac{{err}^2}{F^2F_{var}} \Big)^2 + \Big( \sqrt{\frac{{err}^2}{N}}. \frac{1}{F} \Big)^2 } 	
\end{equation}
where, {\it N} is the number of data points in the lightcurve.
The fractional variability amplitude ({\it F$_{var}$}) estimated during the period of June 26, 2019 to May 31, 2020 (MJD 58660 - 59000) is 0.72$\pm$0.01. This suggests the source showed a significantly high variability  $\sim$ 72\% in $\gamma$-ray and this is discussed further in \S 3.6. 

To characterize the variability of the source in the flaring states, we estimate the variability time scales corresponding to 
F1 and F2 states. The flux doubling time during which the flux changes by a factor of two between consecutive time interval is
expressed as \citep{Zhang_1999}
\begin{equation}
\centering
 t_{d} = \bigg|\frac{( f_{1} + f_{2})(t_2 - t_1)}{2( f_{2} - f_{1})}\bigg|
\end{equation}
where, {\it t$_1$} and {\it t$_2$} are the two consecutive times with fluxes {\it f$_{1}$}, and  {\it f$_{2}$}. 
The shortest variability time (t$_{\rm var}$) then corresponds to the least value of $t_d$ estimated from all the consecutive
time intervals from the entire lightcurve. 
We noted the fastest $\gamma$-ray variability time (t$_{\rm var}$) during flare-1 state
is, 2.059$\pm$0.58 days with the flux values 2.78$\pm$0.89 and 8.04$\pm$1.29 ($\times10^{-7} ph\,cm^{-2}\, s^{-1}$) 
on November 2, 2019 (MJD 58789) and November 4, 2019 (MJD 58791). While for flare-2 state, it is found to be 2.59$\pm$0.79 days with the flux values 
4.26$\pm$1.02 and 9.62$\pm$1.32 ($\times10^{-7} ph\,cm^{-2}\,s^{-1}$) measured during April 8, 2020 (MJD 58947) and April 10, 2020 (MJD 58949), respectively. The obtained t$_{\rm var}$ is close to the binning time, and this is discussed in \S 5.

The flux variability timescale can be used to constrain the size and the location of the emission region. 
Rapid variability observed in $\gamma$-ray indicates the size of the emission region to be of few Schwarzschild radius 
of the central black hole and hence should be located close to the central engine often within the broad-line 
region (BLR). However, the broadband spectral modelling of the FSRQs suggests the $\gamma$-ray emission can be well explained
when the emission region is located outside the BLR (\citealt{2012MNRAS.419.1660S, 2017MNRAS.470.3283S}). 
In this work, we found the fast variability timescale to be of the order of 2 days. This can be used to constrain the 
size of the emission region as R $\lesssim$ c t$_{\rm var} \delta/(1+z)$ where, $\delta$ 
is the jet Doppler factor and c is the speed of light. For $\delta \sim 10$ (\citealt{10.1093/mnras/stx002}), R is estimated as $ \lesssim$ 2.6 $\times 10^{16}$ cm. Similarly, the distance of the emission region can be estimated as d $\sim$ 2 c t$_{var}$ $\delta^2$/(1+z) $\sim$ 5.2$\times 10^{17}$ cm (\citealt{Abdo_2011}). These results are discussed in \S 5.


\begin{table}
\centering
	\caption{Time period for the sub-flares in F1 and F2 states.}
	\begin{tabular}{l c }
		\hline
		     	States		& MJD\\
		\hline
		
$F1_{a}$  & 58780 - 58813\\
$F1_{b}$  & 58813 - 58840\\
$F1_{c}$  & 58840 - 58857\\
$F2_{a}$  & 58857 - 58888\\
$F2_{b}$  & 58888 - 58935\\
$F2_{c}$ & 58935 - 58963\\
$F2_{d}$ & 58963 - 59000\\

	\hline 
	\end{tabular}
	\label{tab:sub-flares}
	
\end{table}

\begin{table*}
	\centering
	\caption{Analysis results for $\gamma$-ray SED, in various flux states.}  
	\label{partable}
 \begin{tabular}{c c c c c c c}
 \hline
		\hline
 States & F$_{0.1-300 \rm{GeV}}$ & Power Law & & & TS & TS$_{curve}$ \\
  & (10$^{-7}$ ph\,cm$^{-2}$\,s$^{-1}$) & $\Gamma$ & & && \\
  \hline
 $F1_{a}$ & 5.8$\pm$0.36&  -2.22$\pm$0.04 & -- & --& 2386.37 & -- \\ 
 $F1_{b}$ & 9.1$\pm$0.28&  -2.14$\pm$0.03 & -- & -- & 5353.15&-- \\
 $F1_{c}$ & 7.6$\pm$0.38&  -2.25$\pm$0.04 & -- & -- & 1915.97&-- \\ 
 \hline
 $F2_{a}$ & 8.3$\pm$0.30&  -2.21$\pm$0.03 & -- & --& 3996.26 & -- \\
 $F2_{b}$ & 10.0$\pm$0.26&  -2.19$\pm$0.02 & -- & --& 7112.10 & -- \\
 $F2_{c}$ & 6.5$\pm$0.27&  -2.22$\pm$0.04 & -- & --& 2599.61 & -- \\
$F2_{d}$ & 6.7$\pm$0.24&  -2.24$\pm$0.03 & -- & -- & 3636.70&-- \\
\hline
\hline
    &&  Log-Parabola & \\
    & & $\alpha$ & $\beta$ &  & \\
  \hline 
  $F1_{a}$ & 5.6$\pm$0.27& 2.10$\pm$0.06 & 0.07$\pm$0.03 & --& 2396.46 & 10.09 \\
  $F1_{b}$ & 8.8$\pm$0.29& 1.93$\pm$0.07 & 0.05$\pm$0.02 & -- & 5369.20&16.05 \\
 $F1_{c}$ & 7.2$\pm$0.39& 2.02$\pm$0.09 & 0.11$\pm$0.03 & -- & 1928.65&12.68 \\
  \hline
$F2_{a}$ & 7.9$\pm$0.32& 2.07$\pm$0.05 & 0.08$\pm$0.02 & --& 3993.41 & -2.85 \\
$F2_{b}$ & 9.6$\pm$0.27& 2.06$\pm$0.04 & 0.07$\pm$0.02 & --&7143.33 & 31.23 \\
$F2_{c}$ & 6.2$\pm$0.28& 2.02$\pm$0.08 & 0.11$\pm$0.03 & --& 2620.01 & 20.40 \\ 
$F2_{d}$ & 6.5$\pm$0.25& 2.15$\pm$0.05 & 0.06$\pm$0.02 & -- & 3656.53&19.83 \\ 
 \hline  
  \hline  
   &&  Broken PowerLaw & & E$_{break}$&   \\
   & & $\Gamma_1$ & $\Gamma_2$ & [GeV] &  \\
 \hline
 $F1_{a}$ & 5.7$\pm$0.27& -2.14$\pm$0.06 & -2.39$\pm$0.09 & 1.01$\pm$0.21 & 2391.14&4.77 \\
 $F1_{b}$ & 8.8$\pm$0.29& -2.04$\pm$0.04 & -2.32$\pm$0.07 & 0.91$\pm$0.14 & 5371.34&18.19 \\
 $F1_{c}$ & 7.3$\pm$0.39& -2.08$\pm$0.07 & -2.69$\pm$0.19 & 1.02$\pm$0.37 & 1927.36&11.39 \\
 \hline
 $F2_{a}$ & 8.0$\pm$0.47& -2.06$\pm$0.08 & -2.53$\pm$0.13 & 1.02$\pm$0.18 & 3995.02 & -1.24 \\
 $F2_{b}$ & 9.7$\pm$0.26& -2.08$\pm$0.04 & -2.47$\pm$0.08 & 1.03$\pm$0.18 & 7138.39 &26.29 \\
 $F2_{c}$ & 6.2$\pm$0.28& -2.04$\pm$0.07 & -2.62$\pm$0.16 & 0.99$\pm$0.35 &2619.60 &19.99 \\
 $F2_{d}$ & 6.5$\pm$0.25& -2.13$\pm$0.06 & -2.44$\pm$0.11 & 0.86$\pm$0.47 & 3655.38&18.68 \\
 \hline 
 \end{tabular}
\label{tab:gammased_fitparams}
\end{table*}
\begin{table}
	\caption{Fractional variability ({\it $F_{var}$}) in different wavebands.}
	\begin{tabular}{l c }
		\hline
		Waveband     			& {\it $F_{var}$} (MJD 58660 - 59000)\\
		\hline
		
{\it Fermi}-LAT (0.1-300GeV) &  $ 0.72 \pm  0.01 $\\
\emph{Swift}-XRT (0.3-10keV) & $ 0.27 \pm 0.05 $\\
UVOT band-B  & $ 0.39 \pm 0.01 $\\
UVOT band-U  & $ 0.33 \pm 0.01 $\\
UVOT band-V  & $ 0.37 \pm 0.02 $ \\
UVOT band-W1  & $ 0.32 \pm 0.01 $\\
UVOT band-W2  & $ 0.34 \pm 0.01 $\\
UVOT band-M2 & $ 0.31 \pm 0.01 $\\

	\hline 
	\end{tabular}
	\label{tab:fvar}
\end{table}
\begin{table}
	\caption{Details of the \emph{Swift} observations during Q, F1$_{b}$ and F2$_{b}$ states}
	\begin{tabular}{l c c c}
		\hline
		  	States	& Facilities	& Obs-ID & Exposure (ks)\\
		\hline
Q & \emph{Swift}-XRT/UVOT & 00041512018 & 1.5 \\ 
\hline
F1$_{b}$ & \emph{Swift}-XRT/UVOT & 00041512021 & 1.2 \\
&\emph{Swift}-XRT/UVOT & 00041512023 & 1.8 \\
&\emph{Swift}-XRT/UVOT & 00041512025 & 1.6 \\
&\emph{Swift}-XRT/UVOT & 00041512026 & 1.6 \\
&\emph{Swift}-XRT/UVOT & 00041512027 & 1.1 \\
\hline
F2$_{b}$ & \emph{Swift}-XRT/UVOT & 00035002131 & 1.6 \\
&\emph{Swift}-XRT/UVOT & 00035002132 & 1.9 \\
&\emph{Swift}-XRT/UVOT & 00035002133 & 1.9 \\
&\emph{Swift}-XRT/UVOT & 00035002134 & 2.2 \\
&\emph{Swift}-XRT/UVOT & 00035002135 & 1.2 \\

	\hline 
	\end{tabular}
	\label{tab:obs}
\end{table}

\begin{table*}
	\centering
\caption{Details of the best fit parameters obtained by fitting the broadband spectrum of states Q, F1$_{b}$, and F2$_{b}$ with the one-zone SED emission model (\citealt{2018RAA....18...35S}) added as a local model in \texttt{XSPEC}.  In the table, column 1: the flux states of the source, column 2: location of the seed target photons for EC process, column 3 $\&$ 4: particle spectral indices before and after the break energy (p1 and p2), column 5: Lorentz factor corresponds to the break energy ($\gamma_{b}$), column 6: bulk Lorentz factor ($\Gamma$), column 7: Magnetic field (B), column 8: $\chi^{2}$/degrees of freedom, column 9: Doppler factor ($\delta$), column 10: Jet kinetic power (P$_{jet}$) in logarithmic scale in units  of erg s$^{-1}$, column 11: Total radiated power (P$_{rad}$) in logarithmic scale in units of erg s$^{-1}$. The subscript and superscript on the parameter values denote the lower and upper bound errors respectively, while `--' indicates the lower/upper bound error is not constrained. The details of parameters that kept frozen during the fitting of the spectrum of Q, F1$_{b}$, and F2$_{b}$ states are discussed in section 4.}  
	
 \begin{tabular}{c c c c c c c c c c c}
 \hline
		\hline
	&  & Parameters & & & & & & Properties & &  \\
		\hline
		
 States & Seed & p1 & p2 & $\gamma_{b}$ & $\Gamma$ & B & $\chi^{2}$/dof &  $\delta$ & P$_{jet}$ & P$_{rad}$ \\ 
 & photons &  & & & &  &  &  &  &  \\ \\
 
Q & IR & $2.69_{-0.74}^{0.23}$ &  $6.63_{-2.37}^{--}$ & ${5.27\times10^{3}}_{-4.38\times10^{3}}^{3.45\times10^{3}}$ & $8.00_{-1.71}^{--}$ & $1.38_{-0.06}^{0.06}$ & 13.93/12 & 14.79 & 45.69 & 43.69  \\ \\

F1$_{b}$ & IR + BLR & $2.42_{-0.26}^{0.14}$ &  $7.49_{-2.97}^{--}$ & ${4.77\times10^{3}}_{-1.75\times10^{3}}^{0.74\times10^{3}}$ & $7.15_{-0.52}^{0.44}$ & $1.59_{-0.03}^{0.05}$ & 11.97/15 & 13.40 & 45.61 & 44.05  \\ \\

F2$_{b}$ & IR + BLR & $1.76_{-0.28}^{0.26}$ &  $5.56_{-0.85}^{--}$ & ${3.57\times10^{3}}_{-0.77\times10^{3}}^{1.40\times10^{3}}$ & $6.14_{-0.39}^{0.52}$ & $1.58_{-0.11}^{0.11}$ & 19.54/16 & 11.67 & 45.20 & 44.43 \\ \\

 \hline 
 \end{tabular}
\label{tab:broadbandsed_fitparams}
\end{table*}

\begin{figure*}
    \centering
    \includegraphics[scale=0.25]{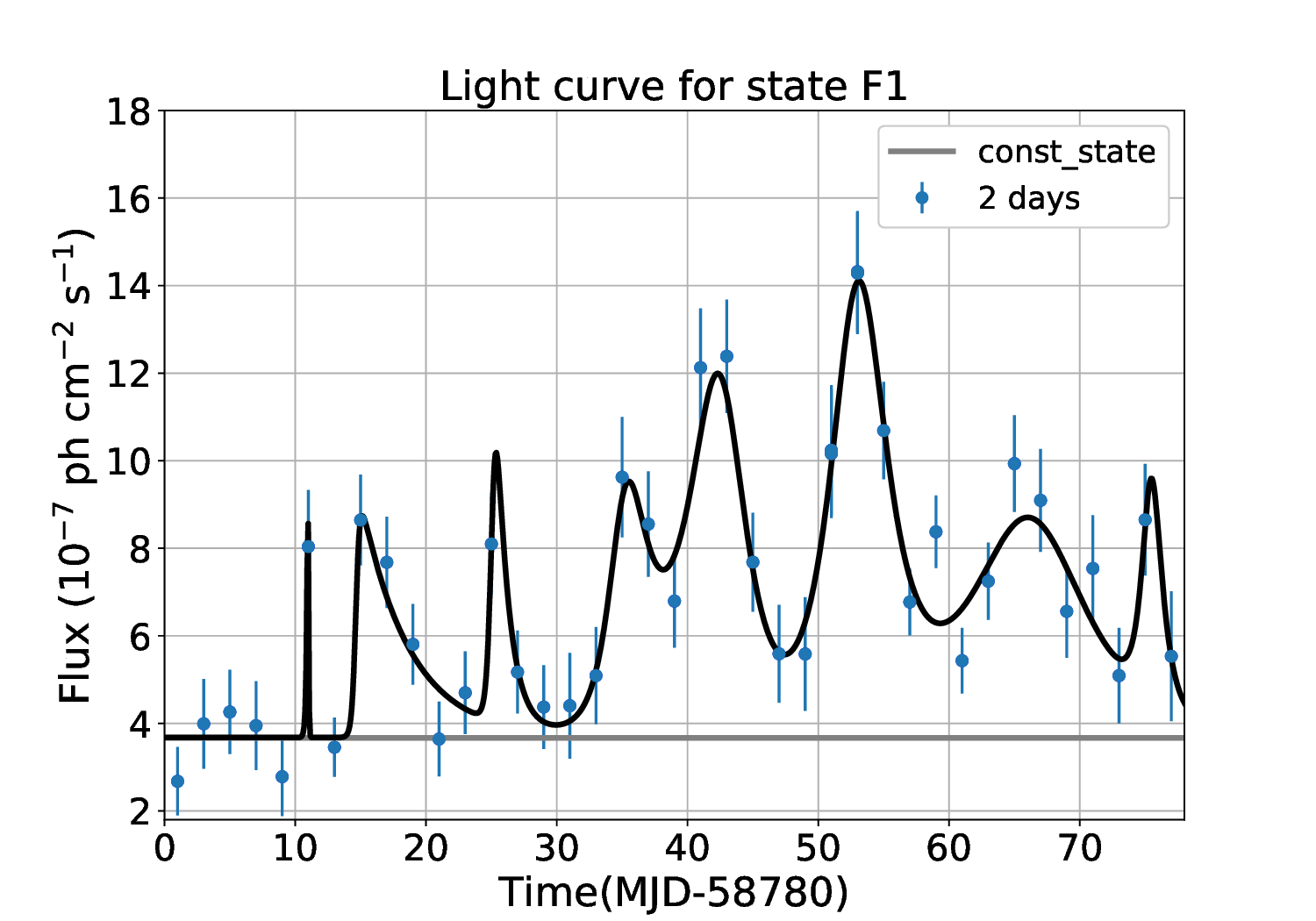}
    \includegraphics[scale=0.25]{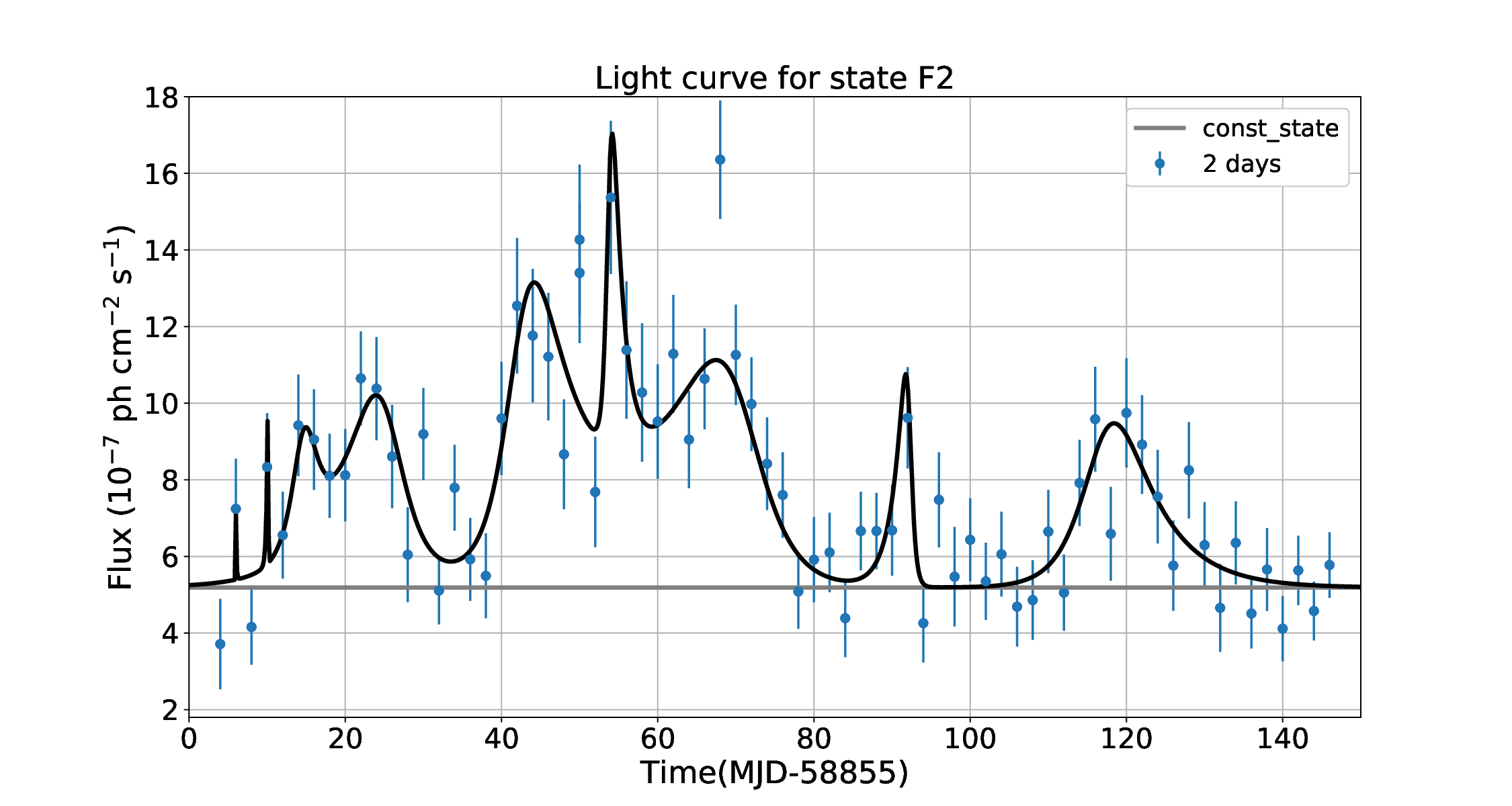}
    \includegraphics[scale=0.38]{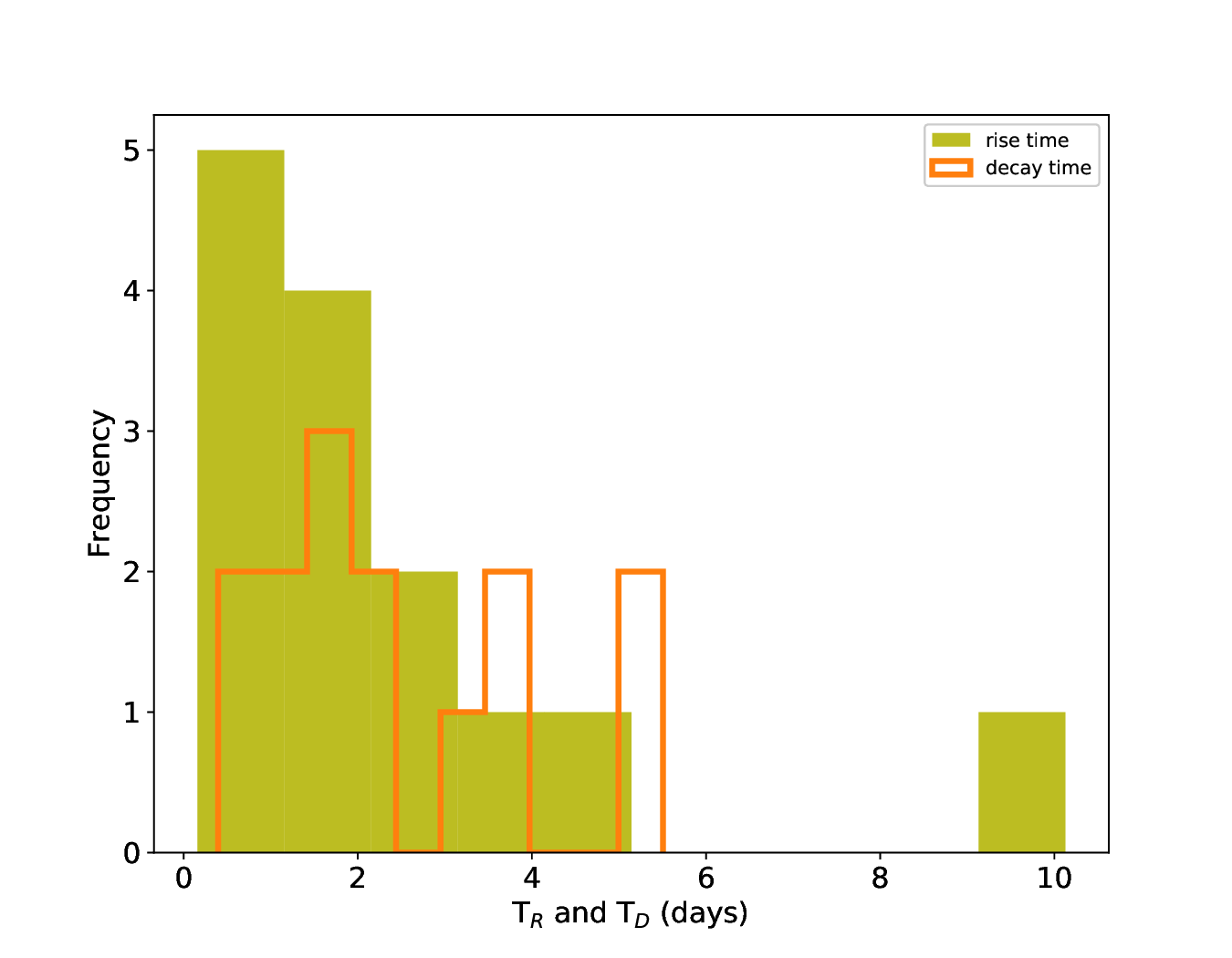}
    \caption{The upper two panels represent the states \texttt{F1 \& F2} fitted with sum of exponential. The lower panel shows the distribution of rise and decay time of the peaks observed during both the states.}
    \label{fig:rise-decay}
\end{figure*}

\begin{figure*}
	 \includegraphics[scale=0.32]{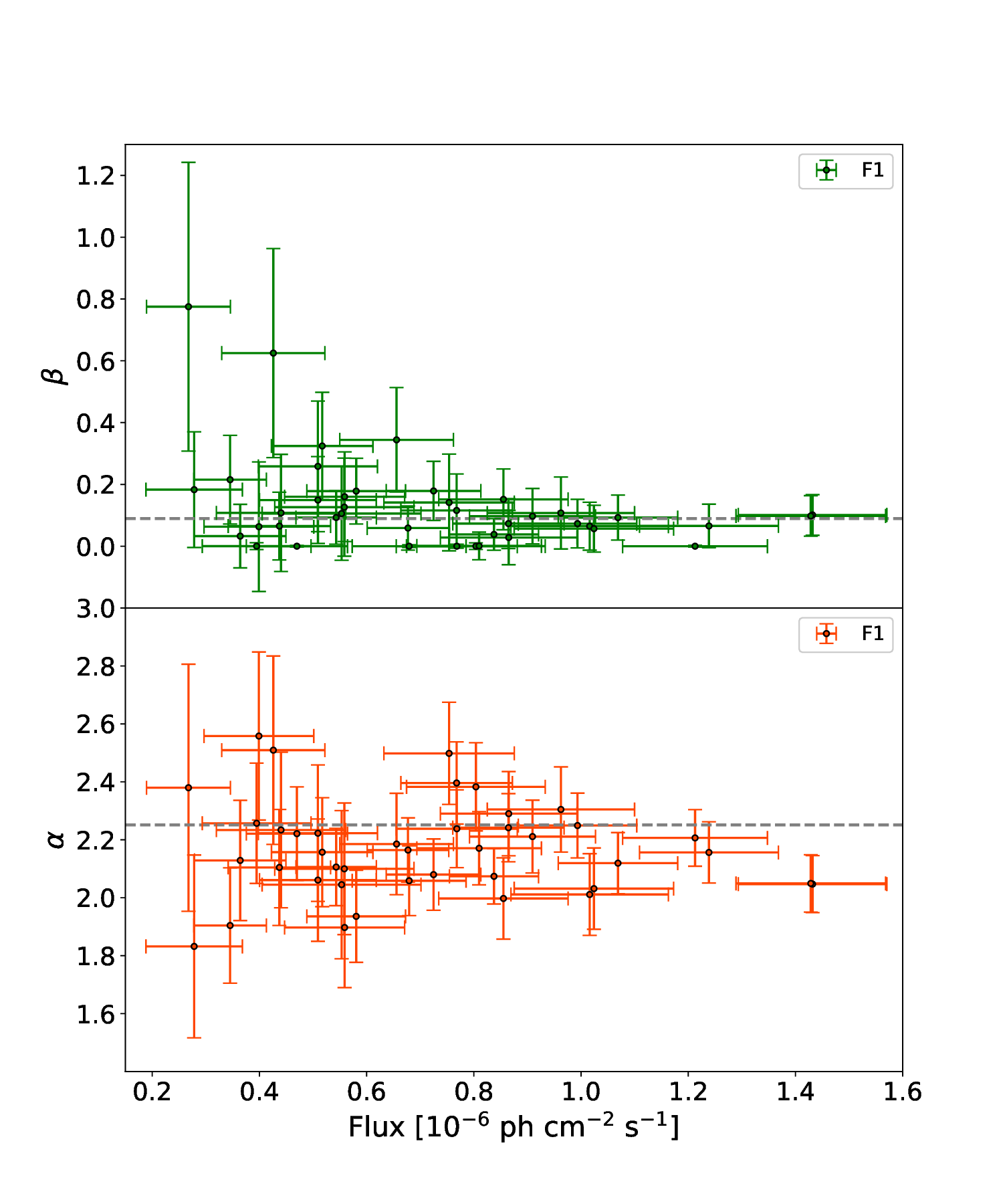}\label{sed-f1a}
       \qquad
       \includegraphics[scale=0.32]{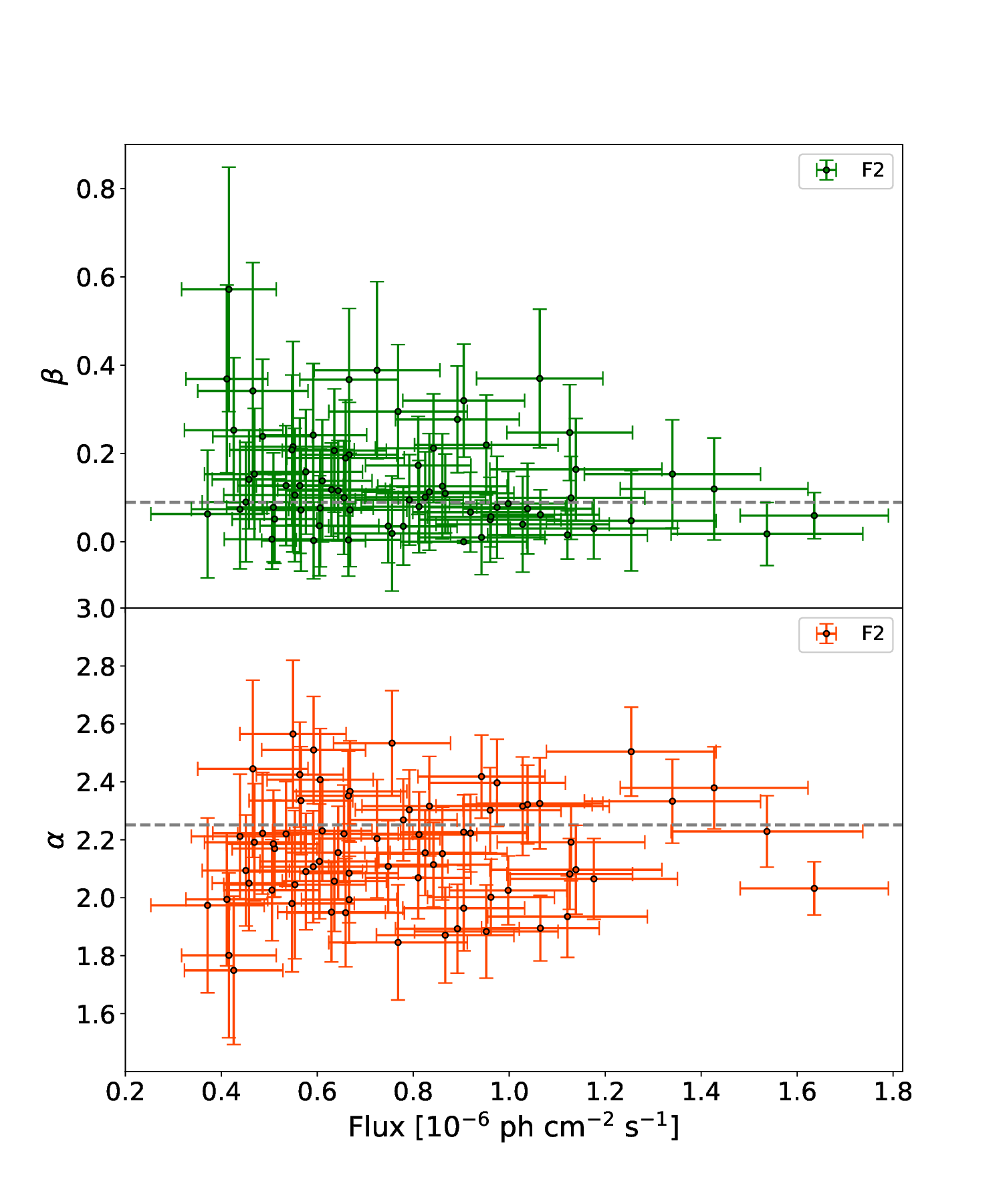}\label{sed-f1b}

       \caption{Scatter plots of the 2-days binned $\gamma$-ray flux (100 MeV - 300 GeV) against spectral index ($\alpha$) and the curature index ($\beta$). Left side top $\&$ bottom plots: Flux ( $\times 10^{-6}$ ph\,cm$^{-2}$\,s$^{-1}$) versus $\beta$ and Flux ($\times 10^{-6}$ ph\,cm$^{-2}$\,s$^{-1}$) versus $\alpha$ correspond to F1 state. Right side top $\&$ bottom plots: Flux ($\times 10^{-6}$ ph\,cm$^{-2}$\,s$^{-1}$) versus $\beta$ and Flux ($\times 10^{-6}$ ph\,cm$^{-2}$\,s$^{-1}$) versus $\alpha$ correspond to F2 state, respectively. }

      \label{fig:spec_variation}
      \end{figure*}

\begin{figure*}
    \centering
    \includegraphics[scale=0.27]{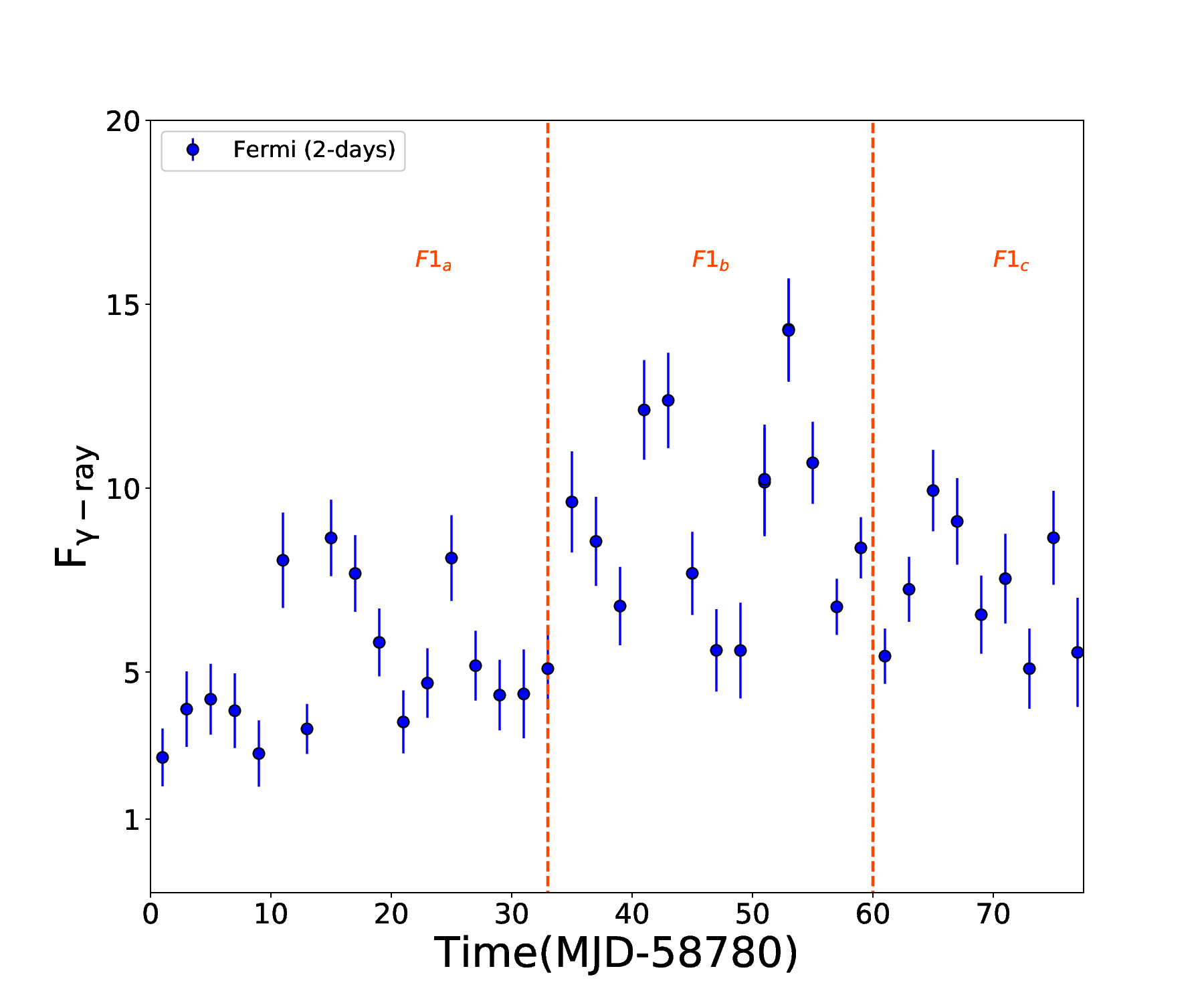}
    \includegraphics[scale=0.27]{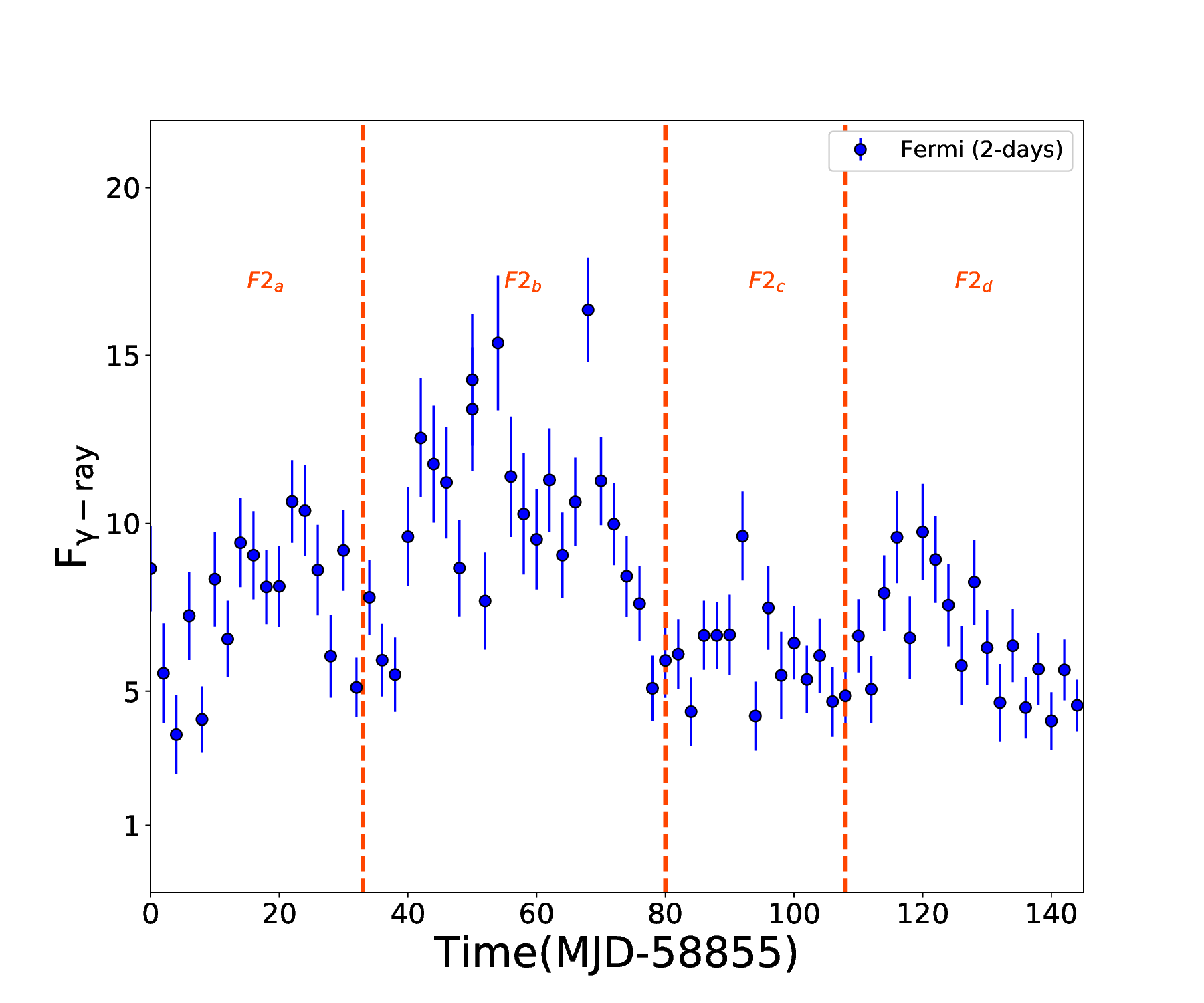}
    \caption{The $\gamma$-ray lightcurves for multiple states in F1 (left plot) and F2 (right plot).}
    \label{fig:F1_F2}
\end{figure*}

We also estimated the rise and decay times for the flares F1 and F2 using sum of exponentials of the form
\begin{equation}
   {\rm F(t)} = {\rm C} + 2 {\rm F_0} \left[ \exp{\left(\frac{t_0-t}{T_{R}} \right)} + \exp{\left(\frac{(t-t_0)}{T_{D}}\right)} \right]^{-1}
    \label{eq:sum_of_exp}
\end{equation}
where, {\it T$_R$} and {\it T$_{D}$} are the rise and decay time of a flare, F$_{0}$ is the peak flux observed at time {\it t$_{0}$} and C is the 
constant introduced to obtain the base flux level. The temporal fit to the flares F1 and F2 indicated many sub flares which are shown 
in the upper panel of Figure \ref{fig:rise-decay}. The reduced chi-square values to the fit of F1 and F2 flares are 0.91 and 1.42, respectively. The histogram of rise and decay time obtained through the temporal fit of the 
flares is shown in the lower panel of Figure \ref{fig:rise-decay}. From the histogram, it is evident that most of the sub flares 
have rise time $\sim$ 1 day, and decay time $\sim$ 2 day. There are a few data points that show a sharp change in the flux value. Albeit the sharp changes are fitted by the sum of exponentials, the {\it T$_R$} and {\it T$_{D}$} values are not reliable and hence they are not included in the histogram.

\begin{figure*}
	\centering
	 \includegraphics[scale=0.3]{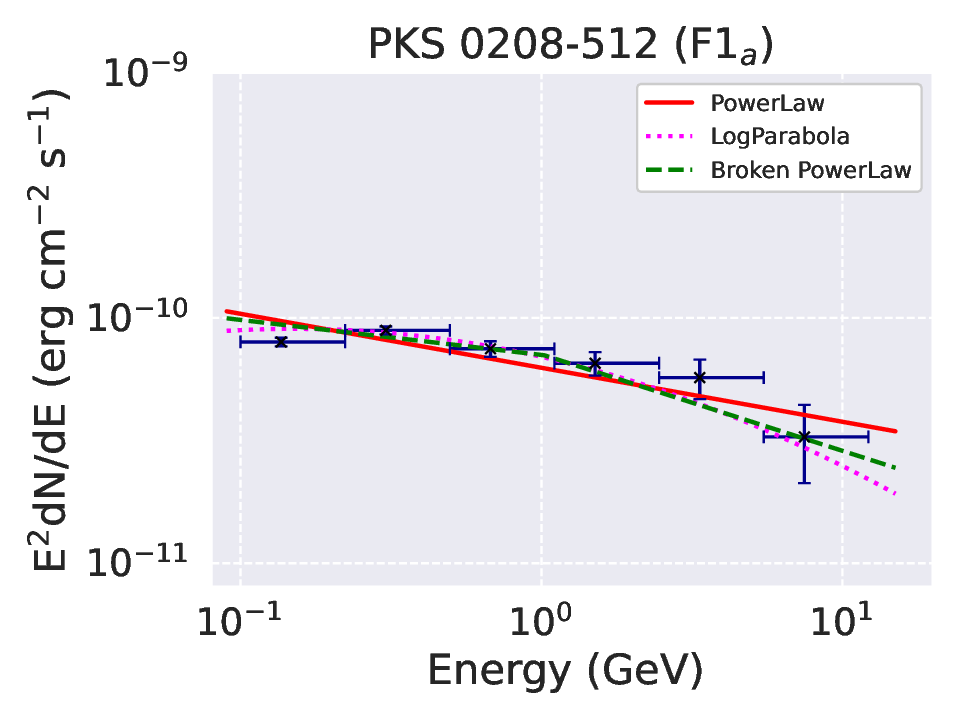}\label{sed-f1a}
       \qquad
       \includegraphics[scale=0.3]{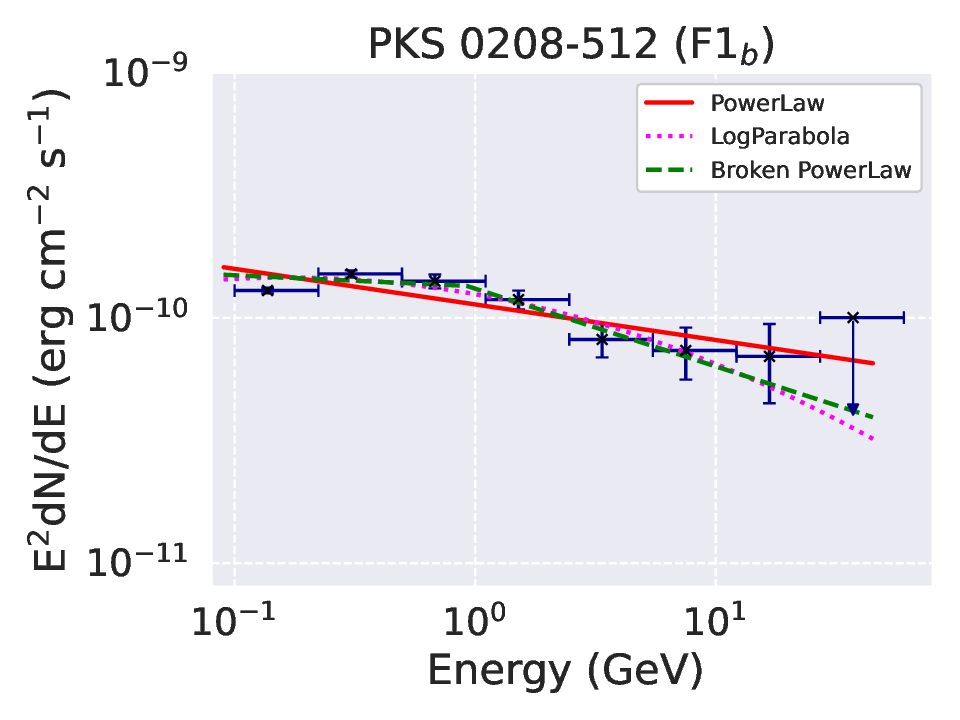}\label{sed-f1b}
       \qquad
       \includegraphics[scale=0.3]{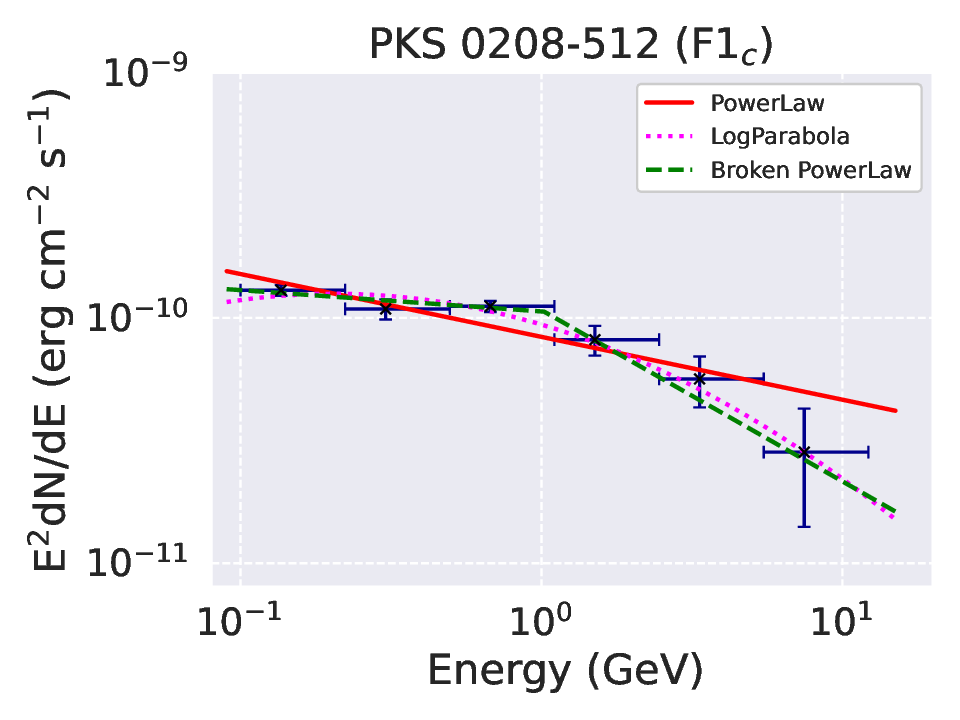}\label{sed-f1c}
       \qquad
       \includegraphics[scale=0.3]{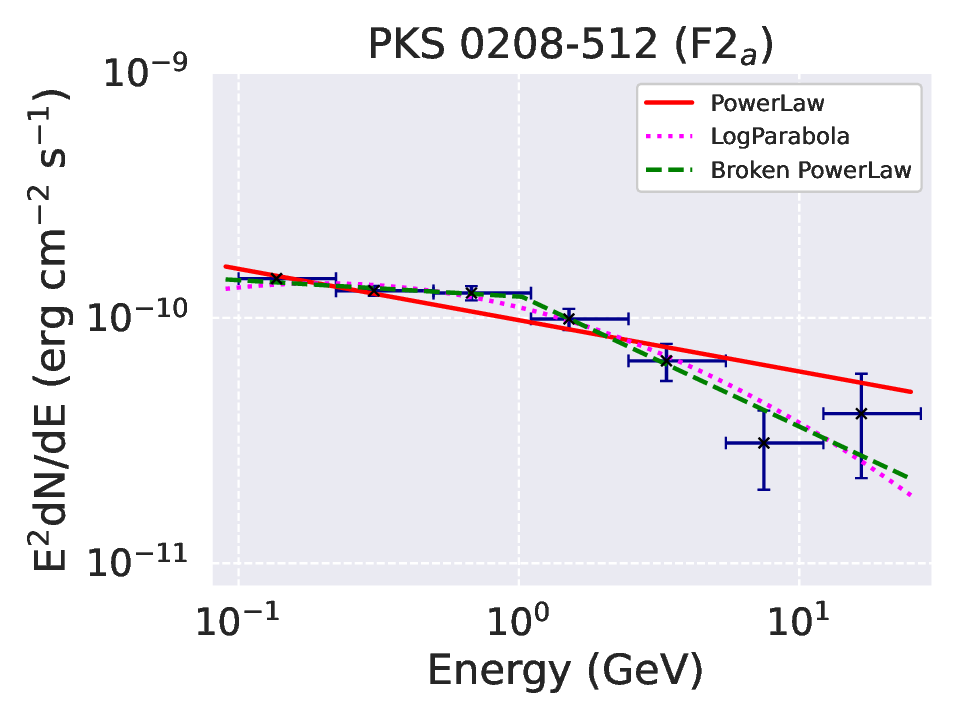}\label{sed-f2a}
       \qquad
       \includegraphics[scale=0.3]{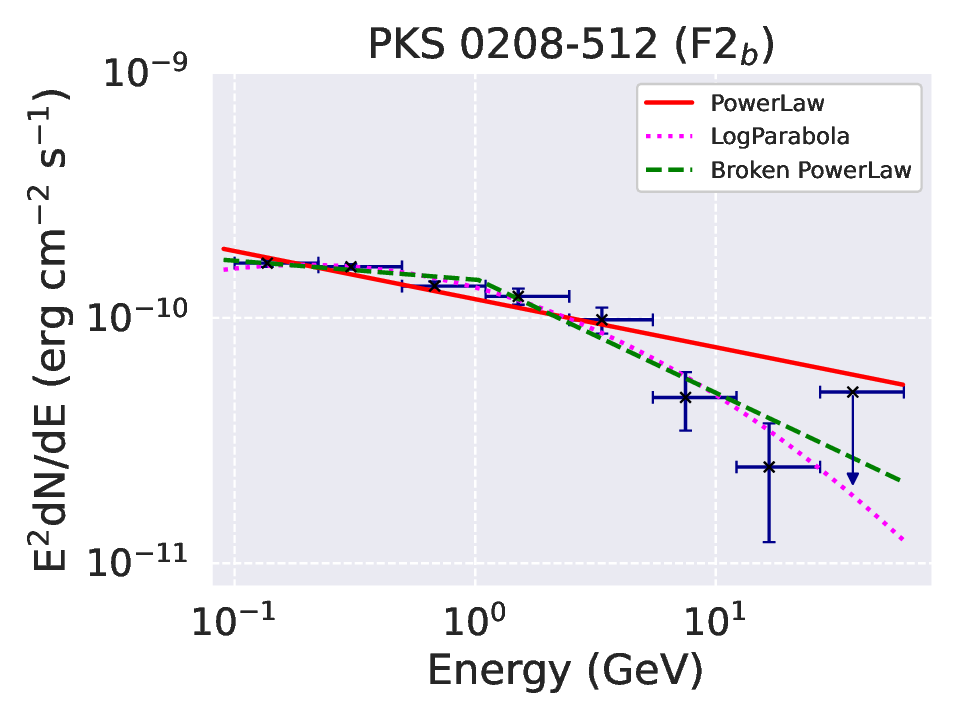}\label{sed-f2b}
       \qquad
       \includegraphics[scale=0.3]{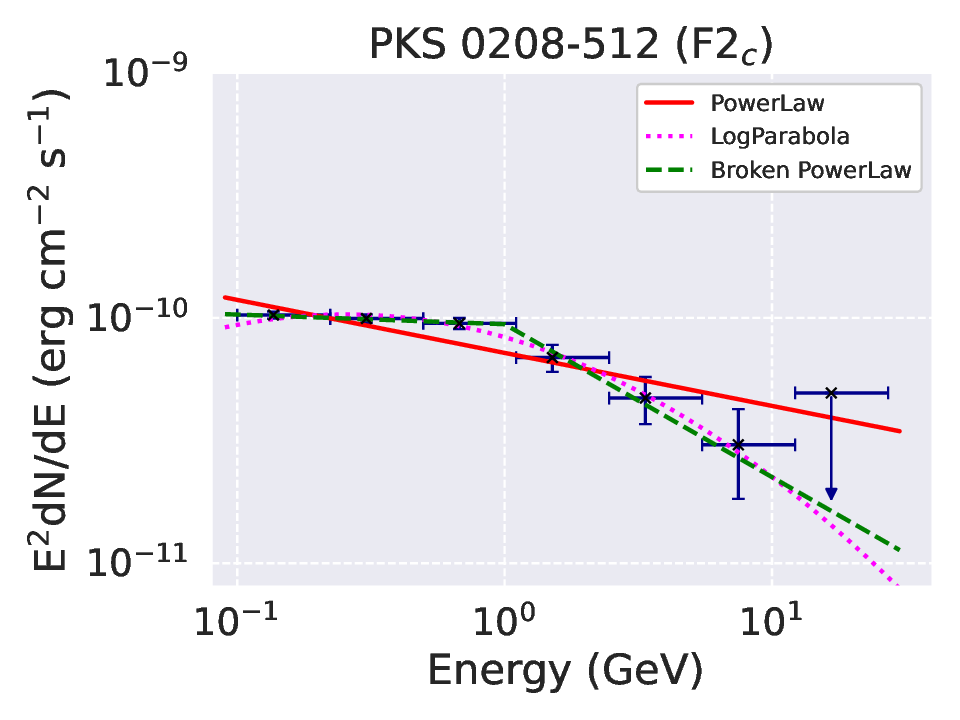}\label{sed-f2c}
       \qquad
       \includegraphics[scale=0.3]{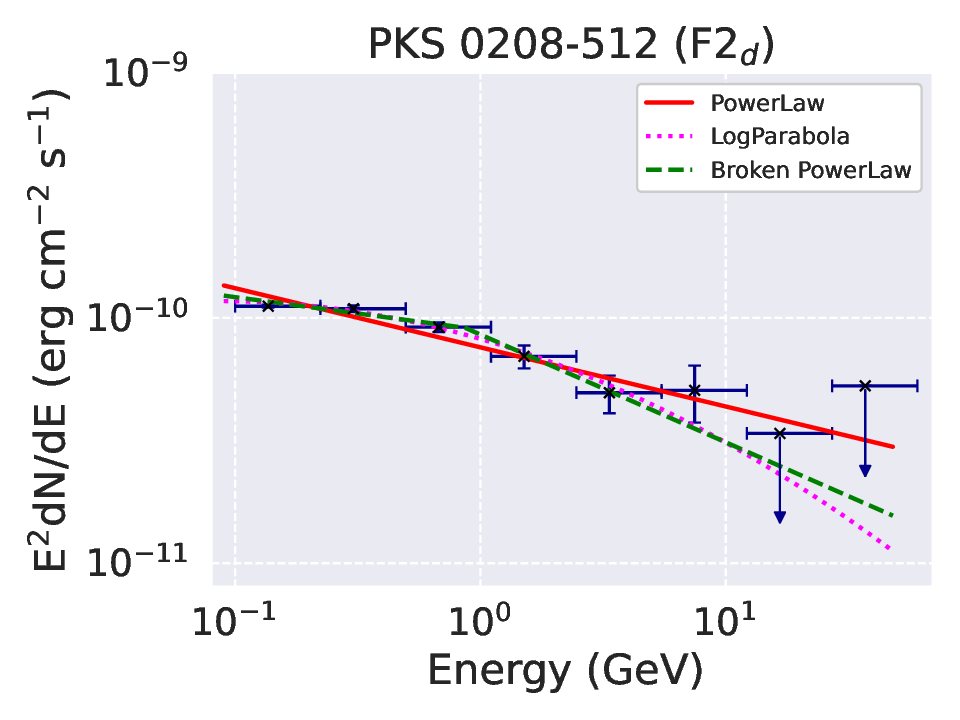}\label{sed-f2d}
       \caption{The $\gamma$-ray Spectral Energy Distribution (SED) for the sub-flare states, fitted with power-law model (red solid line); log-parabola (magenta dotted line) and broken power-law (green dashed line) models, respectively.}
       \label{fig:gamma-sed}
      \end{figure*} 
      

      \begin{figure*}
	\centering
       \includegraphics[scale=0.4]{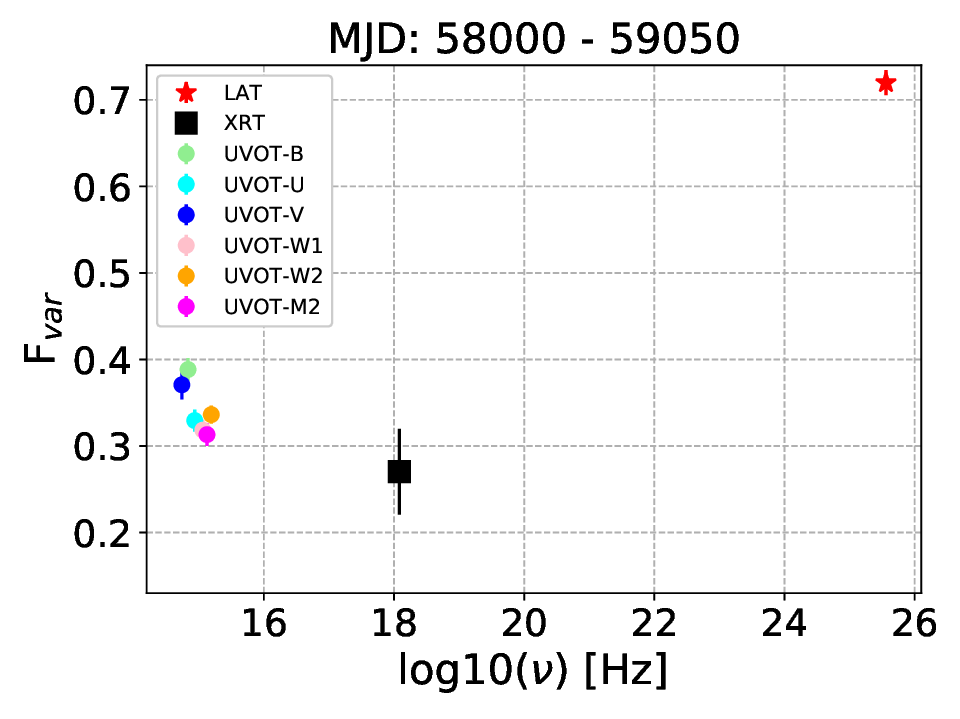}\label{sed-f1b}
       	\caption{Fractional variability vs. frequency plot for each observation (see Figure \ref{fig:mwl_lc}). The numerical values are reported in Table \ref{tab:fvar}. }
	   \label{fig:freq-Fvar}
\end{figure*} 

\begin{figure}
        \centering
        \includegraphics[scale=0.45,angle=-90]{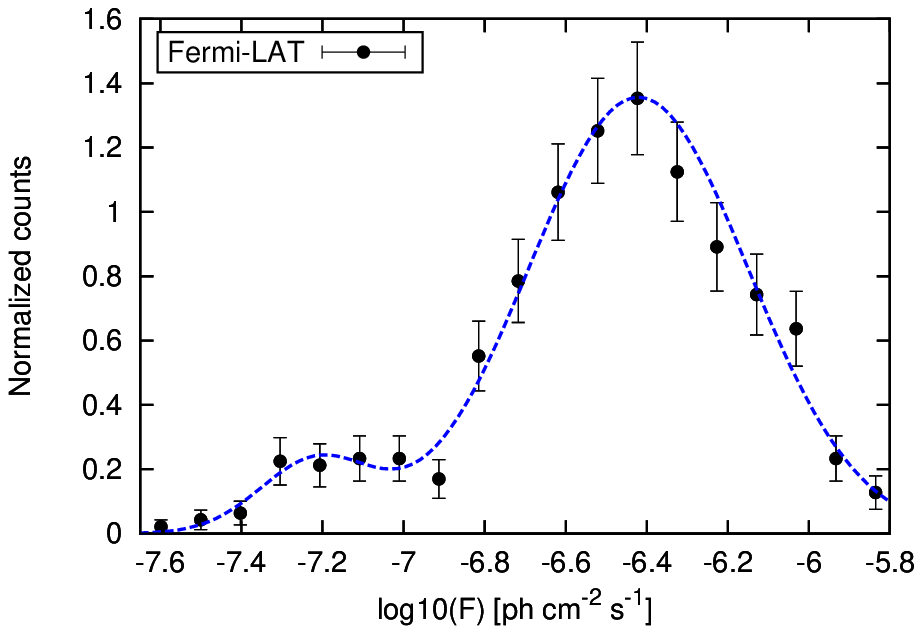}
        \caption{Histogram of the 2-days binned $\gamma$-ray flux of PKS 0208-512. The dashed
blue line corresponds to the best-fitting double lognormal function.}
        
        \label{fig:fermi_hist}
\end{figure}    
      
\begin{figure*}
    \centering
    \includegraphics[scale=0.4,angle=-90]{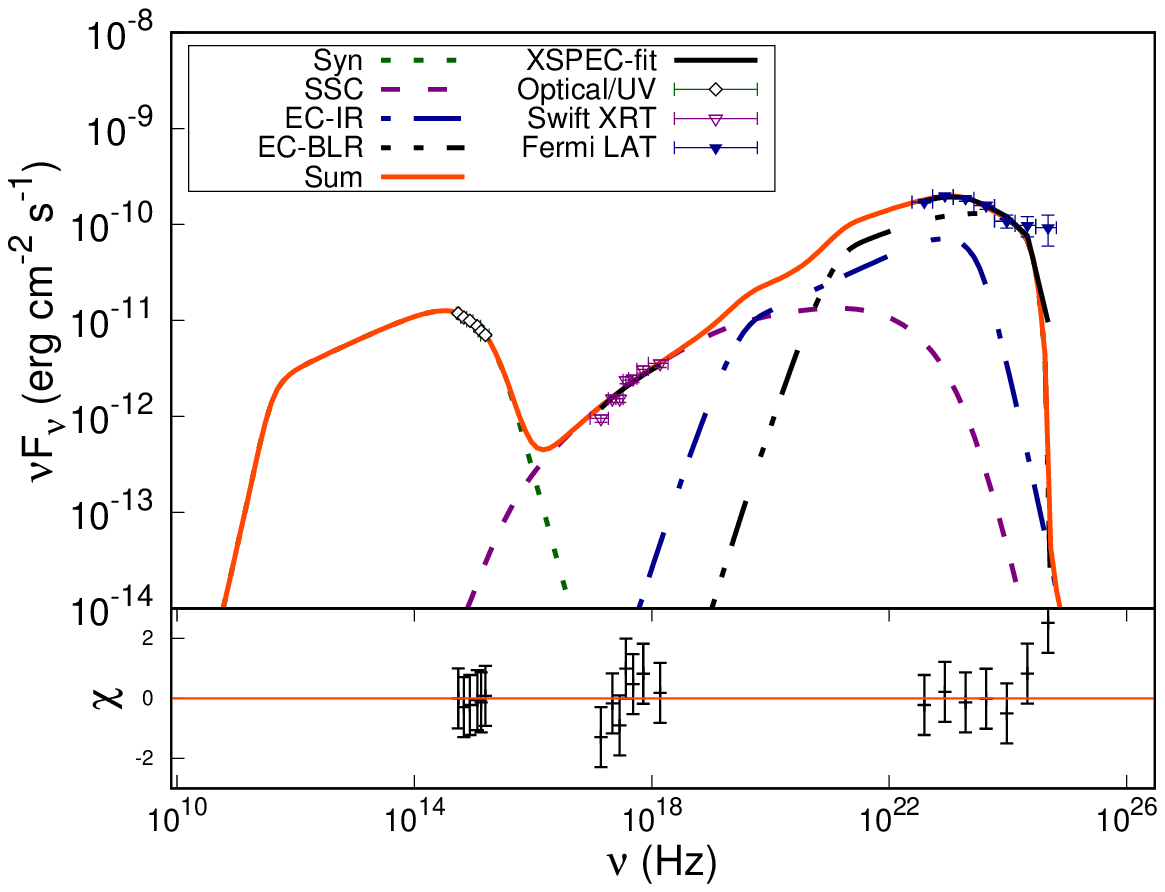}
    \includegraphics[scale=0.4,angle=-90]{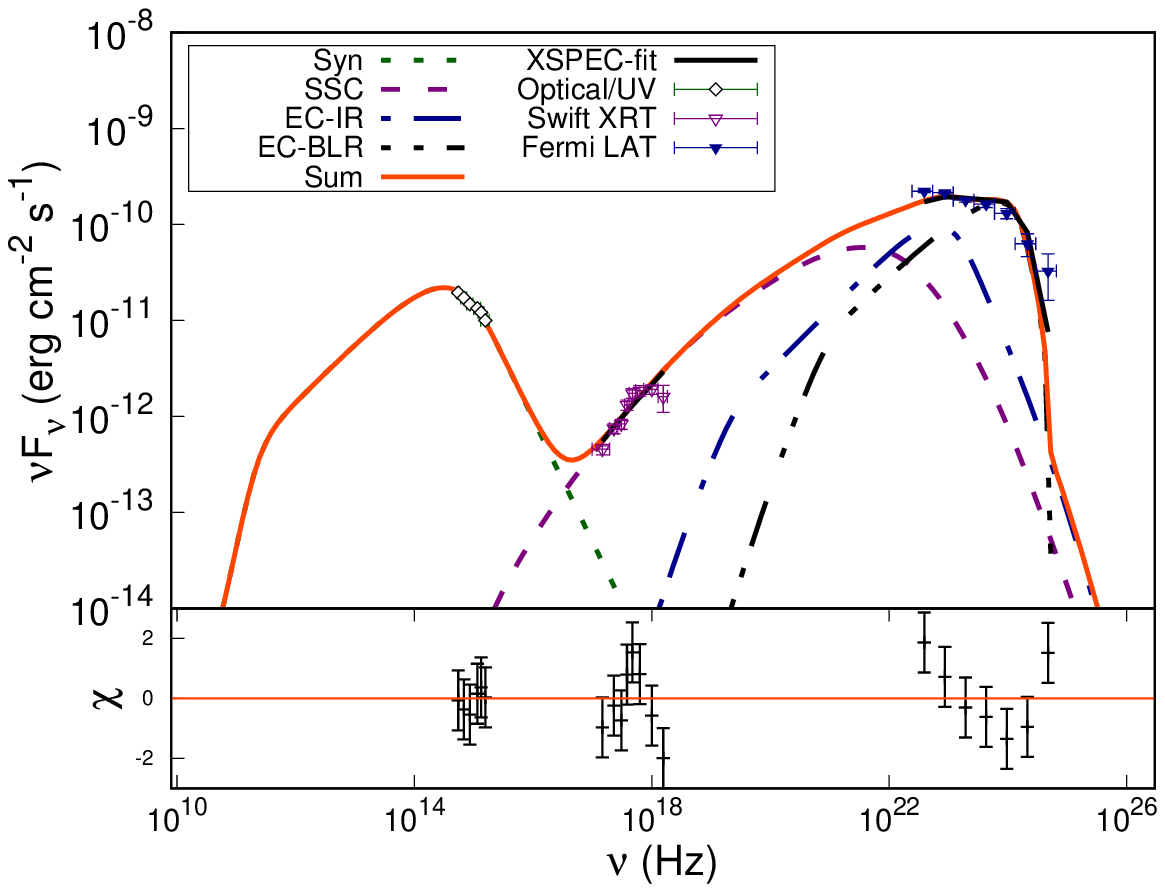}
    \includegraphics[scale=0.4,angle=-90]{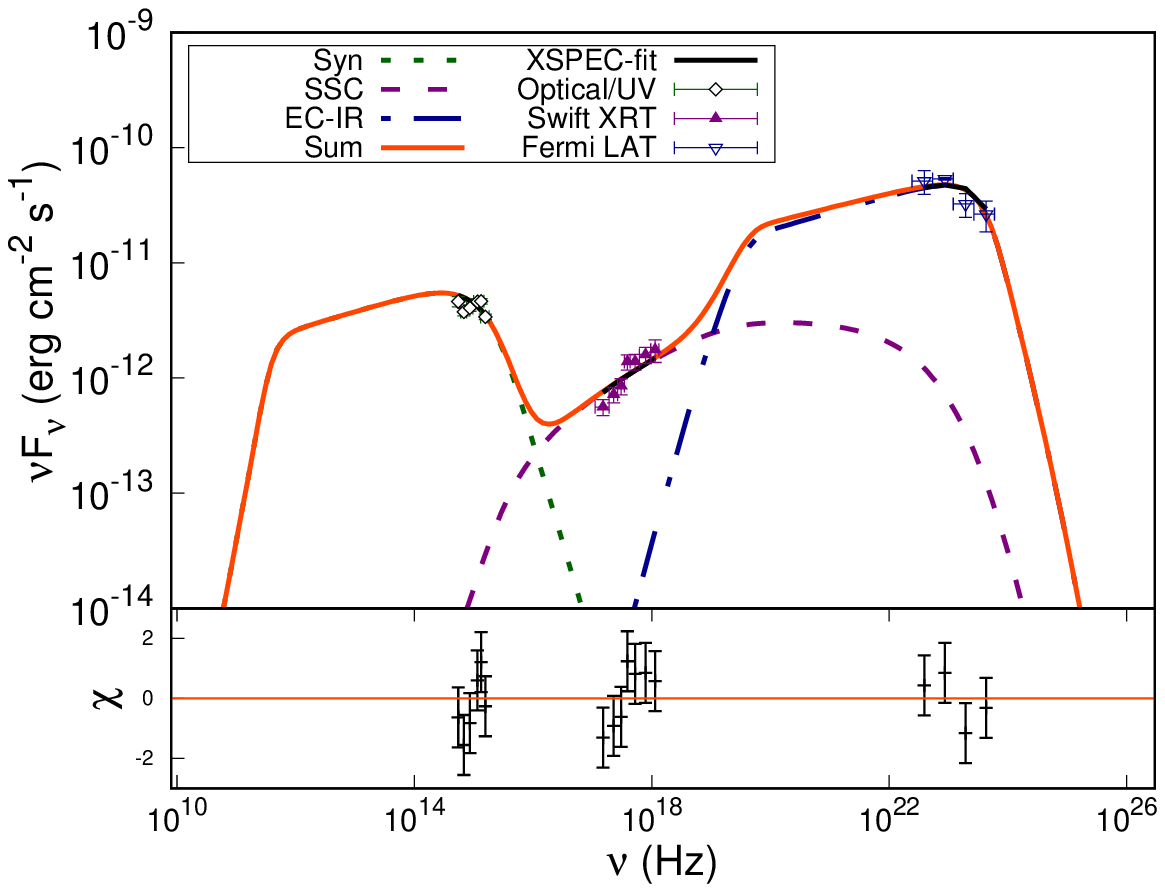}
    \caption{Broadband SED plots of PKS\,0208-512, during flaring states of F1$_{b}$ (top left) and F2$_{b}$ (top right), and quiescent state Q (bottom).}
    \label{fig:multised}
\end{figure*} 

\subsection{$\gamma$-ray spectral variation}

The $\gamma$-ray spectrum of PKS\,0208-512 at the energy range 0.1-300 GeV is curved and better fitted by a log-parabola (4FGL; \citealt{2020ApJS..247...33A}). To understand
the spectral variation of the source during the active state, we performed a Spearman rank correlation study between the parameters defining 
the spectral shape and the 2-days integrated flux. In Figure \ref{fig:spec_variation}, we showed the scatter plots between 2-days binned 
$\gamma$-ray flux (F$_{0.1-300 GeV}$) against the best fit log-parabola parameters $\alpha$ and $\beta$ during the flaring states F1 
(left side top and bottom plots) and F2 (right side top and bottom plots). For the flaring state F1, the Spearman's rank correlation study 
between F$_{0.1-300 GeV}$ and $\alpha$ resulted in rank coefficient $r=-0.07$ with null-hypothesis probability $P=0.66$; while between
F$_{0.1-300 GeV}$ and $\beta$ we get $r= -0.33$ with $P=0.33$. The high value of $P$ suggests the correlation results are inconclusive.
Similarly, for the flaring state F2 the correlation study between F$_{0.1-300 GeV}$ and $\alpha$ resulted in $r= 0.1$ with $P=0.39$, and 
between F$_{0.1-300 GeV}$ and $\beta$ we obtained $r= -0.24$ with $P=0.04$. The low value of P (<0.05) suggests there exists a mild
negative correlation between F$_{0.1-300 GeV}$ and $\beta$ during F2. Though `harder-when-brighter' trend is generally witnessed in case 
of blazars, our correlation study was unable to confirm the same for the case of 
PKS\,0208-512 \citep{2019MNRAS.484.3168S, 2020MNRAS.491.1934K, Prince_2020,2021MNRAS.502.5245P}.
The `harder-when-brighter' or `softer-when-brighter' behavior can be interpreted as an outcome of shift in the Compton spectral peak
when the peak falls in the energy range where {\it Fermi}-LAT is sensitive. For instance, a high energy shift in the Compton peak
during a flare may result in `harder-when-brighter' trend while low energy shift may imitate `softer-when-brighter'. 
However, since no such behavior is observed in our study, it is hard to conclude whether the spectral change is associated
with the flux state of PKS\,0208-512.

Furthermore, we added the default 4FGL $\alpha$ and $\beta$ values i.e., 2.25 and 0.089 in Figure \ref{fig:spec_variation}, as shown by the horizontal grey dashed lines. It can be seen in the Figure that during the states F1 and F2, most of the points are below the 4FGL $\alpha$ value suggesting a harder spectral behavior compared to the 4FGL value. On the contrary, most of the points are marginally above the 4FGL $\beta$ value implying the spectrum is more curved during the flaring state compared to the 4FGL value. 
\subsection{$\gamma$-ray spectral analysis}
 
The flaring states F1 and F2, show multiple sub-flares as shown in Figure \ref{fig:F1_F2}. To examine the spectral changes during these 
sub-flare states, we divided flare F1 into three sub-flares viz. \texttt{F1$_{\rm a}$}, \texttt{F1$_{\rm b}$}, and \texttt{F1$_{\rm c}$} 
(left plot of Figure \ref{fig:F1_F2}), and the flare \texttt{F2} into four sub-flares viz. \texttt{F2$_{\rm a}$}, \texttt{F2$_{\rm b}$}, 
\texttt{F2$_{\rm c}$}, and \texttt{F2$_{\rm d}$} (right plot of Figure \ref{fig:F1_F2}), respectively. Each sub-flare shows rapid rise and decay patterns suggesting a strong flux variability. The time period corresponding to each sub-flare is reported in Table \ref{tab:sub-flares}.
For each sub-flare we generated the $\gamma$-ray spectra with the help of \texttt{likesed.py} code developed by 
the {\it Fermi}-LAT collaboration and available at Fermi science tools webpage\footnote{https://fermi.gsfc.nasa.gov/ssc/data/analysis/user/}. 
We binned the entire energy range 0.1-300 GeV into nine equally spaced logarithmic energy bins. The $\gamma$-ray spectra made of nine data points are further 
fitted with the various spectral models: power law (PL), broken powerlaw (BPL), and log-parabola (LP)(see \S 2.1). These models are defined as 
 
\begin{align}
    N(E)dE &= N_0\times\left(\frac{E}{E_0}\right)^{-\Gamma} dE &- \text{PL} \\
	N(E)dE &= \begin{cases} N_0\times\left(\frac{E}{E_0}\right)^{-\Gamma_1} dE \quad{\rm for}\; E < E_b\\
		N_0\times\left(\frac{E}{E_0}\right)^{-\Gamma_2} dE \quad{\rm for}\; E \geq E_b
    \end{cases} &- \text{BPL} \\
    N(E)dE &= N_0\times\left(\frac{E}{E_0}\right)^{-(\alpha+\beta \, \text{log}(E/E_0))} dE &- \text{LP}
    \label{eq:spectral_models}
\end{align}
where, $N_0$ is the normalization, $E_0$ is the reference energy and $E_b$ is the break energy. The power-law indices are defined as 
$\Gamma$, $\Gamma_1$, and $\Gamma_2$. The maximum log likelihood analysis is used to fit the $\gamma$-ray spectrum.

The fitted $\gamma$-ray spectrum for the sub-flares are shown in Figure \ref{fig:gamma-sed} and their best fit parameters are presented in Table \ref{tab:gammased_fitparams}. Further, we estimated the test statistics of the curvature in the spectrum by calculating the parameter TS$_{curve}$ = 2 [log $\mathcal{L}(LP/BPL)$ - log $\mathcal{L}(PL)$], where $\mathcal{L}$ is the likelihood function (\citealt{2012ApJS..199...31N}). We consider curvature to be significant if TS$_{curve}$ $\geq$ 16 (equivalent to 4-$\sigma$ level). It has been seen that the flaring states F1$_{b}$, F2$_{b}$, F2$_{c}$, F2$_{d}$ exhibit significant curvature and/or breaks for the log-parabola and/or the broken power-law models. The TS$_{curve}$ values are reported in Table \ref{tab:gammased_fitparams}. The implications of these findings are discussed in \S 5.

\subsection{Flux distribution of $\gamma$-ray lightcurve}

Besides the spectral property, another striking feature of blazar lightcurve is the lognormal variability.
The lognormal flux distribution has been observed in many blazars at various energy bands over different time scales (\citealt{HESS_2010, Chevalier_2015, my421, my1011, 2018MNRAS.480L.116S, 2020MNRAS.491.1934K}). The analysis of the flux distribution using {\it Fermi}-LAT lightcurves suggests that {\it Fermi}-blazars largely follow lognormal flux distribution (\citealt{Ackermann_2015, 2018RAA....18..141S, 2018Galax...6..135R,2020ApJ...891..120B, 2021MNRAS.502.5245P}). The observed lognormal distribution in the blazar lightcurves can be 
interpreted as the result of a large collection of mini-jets which are randomly oriented inside the relativistic jet (\citealt{minijet}). Alternatively, \citet{2018MNRAS.480L.116S} showed that a Gaussian perturbation in the particle acceleration time-scale could also 
result in the lognormal flux distribution. 
Besides the single lognormal distribution, flux histograms of some blazars indicate double lognormal feature 
in different wavebands (\citealt{pankaj_ln, 2018RAA....18..141S, 2020MNRAS.491.1934K}). \citet{pankaj_ln}, observed that the source PKS 1510-089 has two lognormal profiles in the flux at near-infrared (NIR), optical, and $\gamma$-ray energies, and suggested that these two profiles are possibly connected to the two different flux states of the source. Further, \citet{2020MNRAS.491.1934K} found 
the double lognormal flux distribution at X-rays for the blazars Mrk 501 and Mrk 421, which can be a result of the double Gaussian distribution in the index. 

We studied the flux distribution of PKS 0208-512 using its $\gamma$-ray lightcurve. 
To ensure good statistics, we consider only the flux points for which $\frac{F}{\Delta{F}}$ > 2 and also with 
test statistic (TS) value > 9 (significance > 3$\sigma$). We use the Anderson Darling (AD) test statistic on the
lightcurve where the null hypothesis checks if the sample is drawn from a normal distribution. 
The null hypothesis probability P < 0.01 would then indicate significant deviation from the normal distribution. Therefore, the P-values < 0.01 for the flux in linear-scale and log-scale would suggest the flux distribution is neither normal nor lognormal.
We found the AD test statistics (r) and P-value for the flux in linear scale are 10.91 and $3.7\times10^{-24}$; 
while for the flux in logarithmic scale, the r-value and P-value are 3.25 and $3.66\times10^{-8}$, respectively. This result suggests the flux distribution is neither normal nor lognormal. To explore this further, we construct the histogram of the logarithm of flux, which shows a double peak 
structure (Figure \ref{fig:fermi_hist}). We fitted the histogram with the double Gaussian probability density 
function (PDF) given by

\begin{equation}
 \rm f(x) = \frac{a}{\sqrt{2\pi \sigma_1^2}} e^\frac{-(x-\mu_1)^2}{2\sigma_1^2} \
       + \frac{(1-a)}{\sqrt{2\pi \sigma_2^2}} e^\frac{-(x-\mu_2)^2}{2\sigma_2^2} 
       \label{eq:dpdf}
\end{equation}
Here, $\rm a$ is the normalization fraction, $\rm \mu_1$ and $\rm \mu_2$ are the centroids of
the distribution with widths $\rm \sigma_1$ and $\rm \sigma_2$, respectively. We obtained the fit statistics  $\chi^{2}$/dof as 13.2/15 and this suggests the histogram of the logarithm of flux follows a double Gaussian PDF, which implies the flux distribution of the source is double log-normal. The best fit parameter values of $\rm \mu_1$, $\rm \sigma_1$ are -7.21$\pm$0.03, 0.14$\pm$0.03; $\rm \mu_2$, $\rm \sigma_2$ are -6.42$\pm$0.01, 0.27$\pm$0.01; with a value as 0.15$\pm$0.02, respectively. The double lognormal profile can be associated with two flux states of the source. 
Examining these flux states in detail demand precise study of the 
spectral evolution of the source. However, poor statistics do not let us to perform such a study at $\gamma$-rays 
and probably similar study at low energies have the potential to confirm this inference.
 
\subsection{Study of \emph{Swift} observations}

Using the \emph{Swift} observations shown in the second, third and fourth panels of Figure \ref{fig:mwl_lc}, we estimated the fractional variability F$_{var}$ (equation \ref{eq:fvar}) in X-ray, optical, and UV bands during June, 2019 - May, 2020 (MJD 58660 - 59000). 
The F$_{var}$ values are depicted in Table \ref{tab:fvar} and in Figure \ref{fig:freq-Fvar} we plot F$_{var}$  against the 
photon frequency. The high F$_{var}$ values of optical/UV and $\gamma$-ray emissions than the X-ray emission may possibly be associated to the energy of relativistic electrons. The optical/UV and $\gamma$-ray  emissions are due to the synchrotron and IC emission of high energy electrons respectively, while the X-ray emission is due to the IC emission of low energy tail of the electron distribution. 
This is evident from the  broadband SED modeling of the source presented in \S 4. However, IC interpretation of the X-ray emission 
is also suggested for many FSRQs through broadband 
SED modelling \citep{2016A&A...590A..61C,2017MNRAS.464..418R, 2021MNRAS.tmp..823S}.

\section{Broadband spectral modelling}
To perform the broadband SED modelling of the source PKS 0208-512 during different flux states, we selected three time segments including one quiescent state (Q state) MJD 58660 - 58676, and two flaring states with MJD 58813 - 58840 (F1$_b$ state) and MJD 58888 - 58935 (F2$_b$ state) from the lightcurve as shown in Figure \ref{fig:mwl_lc}. The selection was made such that each segment have simultaneous $\gamma$-ray, X-ray, UV and optical observations. 
The details of the \emph{Swift} observations for states Q, F1$_{b}$, and F2$_{b}$ are given in Table \ref{tab:obs}.

We carried the broadband SED modelling of the three selected states, using the one-zone leptonic model (\citealt{2018RAA....18...35S}). In this model, we assume a scenario where the non-thermal emission from the blazar jet  originates from a spherical blob of radius `R' moving with a bulk Lorentz factor $\Gamma$ down the jet at an angle $\theta$ with 
respect to the observer's direction. The emission region is permeated with the tangled magnetic field B, 
and populated by a broken power-law electron distribution described as
\begin{align}
\label{eq:broken}
N(\gamma) d\gamma =\left\{
	\begin{array}{ll}
K \gamma^{-p1}d\gamma,&\mbox {~$\gamma_{{\rm min}}<\gamma<\gamma_b$~} \\
K \gamma^{p2-p1}_b \gamma^{-p2}d\gamma,&\mbox {~$\gamma_b<\gamma<\gamma_{{\rm max}}$~} 
\end{array}
\right.
\end{align}
Here, K and $\gamma$ are the normalization and the electron Lorentz factor, while the low and high energy power-law indices are denoted by  p1 $\&$ p2. The $\gamma_{min}$, $\gamma_{max}$ are the minimum, and maximum Lorentz factors of the electron, and $\gamma_{b}$ is the Lorentz factor corresponding to the break energy.
The observed broadband emission from the source is modelled as synchrotron and inverse Compton (IC) emission by this electron distribution. 
The seed photons for the IC process are the synchrotron photons themselves (SSC) and the photon field external to the jet (EC). 
The source of external target photons considered for the SED modelling are the IR photons from the dusty torus (EC/IR) and the 
dominant Lyman-alpha line emission from the broad line region (EC/BLR). The emission from the dusty torus is assumed to be a black body 
with temperature T $\approx$ 1000K (\citealt{2005A&A...437..861S}). For numerical simplicity, the line emission from BLR is modelled as a black body at T$\approx$ 42000 K (\citealt{Peterson_2006}). 
Since the final emissivity is estimated by convolving with a relatively broad particle distribution, this simplification will have a negligible effect on the resultant IC spectrum. The numerical model estimating the synchrotron and IC emissivity and the observed flux 
from the source parameters is added as a local model in \texttt{XSPEC} \citep{1996ASPC..101...17A}, to perform a statistical fit of the SED. The ASCII format of the optical, UV, X-ray, and 
$\gamma$-ray fluxes are converted to \emph{pha} format using \emph{ftflx2xsp} command and the fitting is performed using $\chi^{2}$-statistics implimented in \texttt{XSPEC}.

The SED is mainly governed by the parameters: K, p1, p2, $\gamma_{b}$, B, R, $\Gamma$, $\theta$, temperature T 
and the fraction of the external target
photons up scattered by the IC  process. Clearly, the number of parameters exceeds the information available at four wavebands. Hence, to constrain the parameters, we enforce additional equipartition between particle and the magnetic 
field energy densities. The fitting is performed by first running the model on the observed SED to obtain acceptable residuals. Then
the confidence intervals on parameters are obtained by freezing certain parameters to their best fit value. This is to avoid degeneracy
introduced by the limited available observed information. Besides these, a 15 \% error was evenly added to the data to get
the reduced $\chi^2_{\rm red}<2$. This limit is demanded by \texttt{XSPEC} to obtain the confidence intervals on the fit parameters.
The best fit model and the parameters corresponding to Q, F1$_b$, and F2$_b$ are given in Figure \ref{fig:multised} and Table \ref{tab:broadbandsed_fitparams}. The parameters
which are kept constant for all the epochs are the size of the emission region R = 2.44$\times10^{16}$ cm (consistent with the observed variability time), 
viewing angle 
$\theta \approx$ 2$^\circ$, the minimum and maximum values of particle Lorentz factor ($\gamma_{min}$ and $\gamma_{max}$) = 50 and $9.31\times10^{5}$, 
and the energy density of the up scattered IR photon field as $\approx 4\times10^{-12}$ erg/cm${^3}$. On the other hand, the 
parameters  p1, p2, $\gamma_{b}$, $\Gamma$, and B are let to vary freely.

The $\gamma$-ray spectrum corresponding to the quiescent state can be readily explained by EC/IR mechanism alone; however, for the flaring states a combination of EC/IR and EC/BLR is required to fit the SED successfully. For the latter, the energy density of  the up scattered BLR 
photons is set at $\approx 3\times 10^{-16}$ erg/cm${^3}$. 
If the emission region is closer to the central black hole within BLR then the $\gamma$-ray spectrum will have the 
contribution from both EC/BLR and EC/IR. On the other hand, if the emission region is away from BLR but within the IR
torus then EC/IR will be dominant. Therefore, our broadband spectral modelling suggests during the activity state 
the emission region is close to the central black hole than the quiescent state. This inference; however, may banks upon the assumptions (e.g. equipartition) imposed on the theoretical SED model and the choice of fixed parameters.

\section{Summary and Discussions}
The long-term monitoring of PKS 0208-512 by {\it Fermi}-LAT, together with the simultaneous observations in other wavebands by \emph{Swift}, allow us to study 
the temporal and spectral behavior of the source in detail. The source showed an intense flaring activity during MJD 58780-59000. 
During this period, the source revealed two major outbursts, viz. F1 and F2. We obtained the fastest variability time-scale
during F1 and F2 states as $\sim$ 2 days. However, using a 2-days binned lightcurve, the obtained variability time-scale of the order of 2-days may not be accurate. This may have an impact on the estimation of size and location of the emission region. Therefore, we generated the 1-day binned $\gamma$-ray lightcurve and estimated the fastest variability time of 0.95$\pm$0.35 day in F1-state, the corresponding values of R and d are obtained as $\sim$ 1.2$\times 10^{16}$ cm and $\sim$ 2.5$\times 10^{17}$ cm, respectively. These values do not deviate much from the values obtained from 2-days binned lightcurve and differ only by a factor of $\sim$2. Nevertheless, the larger error on the flux points in the 1-day binned lightcurve hinders us in locating the peaks of flares. Therefore, we considered the 2-days binned lightcurve in order to obtain the temporal characteristics of the source.
Using 2-days variability time, the size and location of the $\gamma$-ray emission region are estimated as $ \lesssim$ 2.6 $\times 10^{16}$ cm and $\sim$ 5.2$\times 10^{17}$ cm, respectively. Considering a black hole of mass $10^{9}$ \(M_\odot\) (\citealt{2011MNRAS.414.2674G}), the derived emission region size is $\sim$ 2 order of magnitude larger than the Schwarzschild radius ($ \sim$ 3.0$\times 10^{14}$ cm). Relatively a longer time variability has been reported by using \emph{EGRET} observations. For instance,\citealt{1995ApJ...440..525V} found the variability time-scale of 8 days and this corresponds to the size of  $\gamma$-ray emission 
as $\sim$ 3.2$\times10^{16}$ cm for $\delta$ = 3.1. The authors considered a black hole of mass 10$^{10}$ \(M_\odot\) and the derived emission region size is $\sim$ 16 times larger than the Schwarzschild radii (1.9$\times10^{15}$ cm). A variability time-scale of 4-days was also reported and this relates to the emission region of size $\sim$ 5$\times10^{15}$ cm for $\delta \sim 1$ (\citealt{article}). However, the constraints obtained from the pair production opacity of the  $\gamma$-rays suggest a lower limit on the Doppler factor as $\delta \gtrsim 3.2$ (\cite{article}). A variability time of 4-days would then imply the distance of the emission region as, d $\sim$ 1.06$\times10^{17}$ cm. For a disk luminosity L$_{d}$ = 18 erg/sec (\citealt{2011MNRAS.414.2674G}), the size of the BLR can be estimated as R$_{BLR}$ $\sim$  4.24$\times10^{17}$ cm. These results imply that the $\gamma$-rays are produced from a region close to 
the black hole.
The multi-wavelength variability study between the fluxes in different wavebands suggests the variability is large in case of $\gamma$-rays and optical/UV compared to that of X-rays. This can be interpreted in terms of the electron energy responsible for the emission at these wavebands. The optical/UV emission corresponds to synchrotron emission from high energy electrons, whereas X-rays can be due to the IC emission from low energy electrons. Similarly, the $\gamma$-ray emission can be due to IC emission from high-energy electrons. This is also evident from the broadband SED modelling of the source.

The multiple peaks observed in the 2-days binned $\gamma$-ray flux lightcurve during states F1 and F2 are fitted 
with sum of exponentials. The result suggests that most of the peaks have characteristic rise and decay times of the order of 1-day.
This finer sub-structures seen in $\gamma$-ray lightcurves of F1 and F2 are further 
divided as F1$_{a}$, F1$_{b}$, F1$_{c}$, F2$_{a}$, F2$_{b}$, F2$_{c}$, F2$_{d}$. The $\gamma$-ray spectral
study of these sub-flares modelled by the PL, LP, and BPL model 
shows that the states F1$_{b}$, F2$_{b}$, F2$_{c}$, F2$_{d}$ revealed a significant curvature   
with TS$_{curve}$ values as 16.05, 31.23, 20.40, and 19.83, respectively. The observed curved spectrum is best fitted by the 
log-parabola (LP) and/or the broken power-law (BPL) functions. This also suggests significant curvature in the underlying electron 
distribution. Such curvatures in the electron distribution can be attributed to the energy dependence of the particle acceleration/escape time scales (\citealt{2004A&A...413..489M, 2018MNRAS.478L.105J, 2018MNRAS.480.2046G}). 
In  \citealt{2004A&A...413..489M}, the authors showed that a curved particle distribution could originate when 
the acceleration probability is energy-dependent. The energy dependence of the escape timescale in the acceleration 
region can also be responsible for the curvature observed in the spectrum  (\citealt{2018MNRAS.478L.105J, 2018MNRAS.480.2046G}). Alternatively, EBL induced pair production absorption can also be a cause for curved $\gamma$-ray spectrum. However, we observed photons below 20 GeV in all the states and the EBL induced absorption optical depth for the redshift of PKS\,0208-512 (z=1.003) at this $\gamma$-ray energies are very minimal (\citealt{2008A&A...487..837F}). Hence it cannot account for the curvature observed in the $\gamma$-ray spectrum.

The broadband SED of PKS\,0208-512 during different flux states have been fitted by a radiative model considering 
synchrotron, SSC, and EC processes. 
We find that the one-zone leptonic model can successfully explain the SED for all these three states. From the best fit parameters,
it appears that no significant variation in the bulk Lorentz factor $\Gamma$ happened during different flux states, while a marginal decrease is noticed during the flare states. However, 
the observed information is not sufficient enough to constrain all the parameters and hence this conclusion cannot be treated
as a robust one. Further, the $\Gamma$ values obtained are less compared to the 
typical value as observed in other FSRQs (\citealt{2017MNRAS.470.3283S, 2019MNRAS.484.3168S}). 
Nevertheless, our broadband SED fitting suggests that the $\gamma$-ray spectra during flaring (F1$_{b}$ $\&$ F2$_{b}$) states demand 
the EC scattering of photons from both BLR and dusty torus. On the other hand, for the quiescent (Q) state, EC scattering of IR photons 
alone is capable to explain the $\gamma$-ray spectrum. This result can be interpreted in terms of the location of the emission region
during different activity states.
For instance, during F1$_{b}$ and F2$_{b}$ states, the emission region is well within BLR and hence the photons from BLR as well as IR torus will be upscattered through EC process. While in the case of Q state, the emission region may lie ahead of the BLR. 
From Table \ref{tab:broadbandsed_fitparams}, it appears $\Gamma$ decreases marginally during the flare states. This may appear to
contradict to our
inference since the emission region is closer to the central black hole during active states and it is expected the $\Gamma$ to be larger.
However, within the estimated confidence intervals we do not find any significant change in $\Gamma$. Additionally, any deviation from the equipartition 
during the flaring state can also modify $\Gamma$ significantly. The flare state is also associated with a marginal increase in magnetic field and the particle 
energy density (through equipartition). This maybe associated with efficient particle acceleration during the flaring states as compared to the 
quiescent state. This is also consistent with the change in power-law index (p1) of the particle distribution where it hardens during the flaring states.
Non-availability of sufficiently high energy spectra (which can probe the high energy tail electron distribution) 
do not let us to obtain the upper bound of p2. However, its quite evident the difference between p1 and p2 is larger than 1/2 and 
hence the origin of the broken power-law may not be strictly associated with the radiative cooling interpretation of a 
power-law electron distribution (\citealt{kardashev}). 

For MeV blazars, X-rays can be due to the EC mechanism as well (in the framework of leptonic model, see
e.g. \citealt{2009MNRAS.397..985G, 2015MNRAS.446.2483S, 2016ApJ...826...76A, 2020ApJ...889..164M}). However, \citealt{2013ApJ...771L..25C} has performed the broadband study of PKS\,0208-512 using one-zone leptonic model, where the optical, X-ray, and $\gamma$-ray emissions can be explained by the synchrotron, SSC, and EC processes. Further, explanation of X-ray emission of FSRQs through EC process demand a significant deviation from the equipartition (\citealt{2012MNRAS.419.1660S, 2017MNRAS.470.3283S}). In the present work also, we find that the SSC and EC interpretation of the X-ray and $\gamma$-ray emission from PKS\,0208-512 do not require any deviation from the equipartition condition. The optical/UV/X-ray observation period considered for the broadband SED modelling does not encompass the selected $\gamma$-ray period fully. However, we believe the average flux obtained during this period do not change the parameters obtained from the broadband fitting significantly unless extreme low/high optical/UV/X-ray fluxes are missed in the time gap. Alternatively, reducing the $\gamma$-ray observation period consistent with the optical/UV/X-ray observation will not deviate the average $\gamma$-ray flux significantly. On the other hand, this may impact in poor statistics due to less number of $\gamma$-ray photons. Hence, the physical parameters obtained through this study largely define the source within the ambit of the model assumptions. Future observations at very high energy by upcoming experiment Cherenkov Telescope Array (\citealt{2019EPJWC.20901038C}) have a better potential to relax some of the underlying assumptions and gain better constraint on the physical parameters.

\section*{Acknowledgements}

We thank the anonymous referee for valuable comments and suggestions. We acknowledge the use of $\gamma$-ray data, software obtained from Fermi Science Support Center (FSSC), and the \emph{Swift}-XRT/UVOT data from NASA's High Energy Astrophysics Science Archive Research Center (HEASARC), service of Goddard Space Flight Center. This work has made use of the XRT Data Analysis
Software (XRTDAS) developed under the responsibility of the ASI Science Data Center (ASDC), Italy. R.K. and R.G. would like to thank CSIR, New Delhi (03(1412)/17/EMR-II) for financial support. R.P. acknowledges the support by the Polish Funding Agency National Science Centre, project 2017/26/A/ST9/00756 (MAESTRO 9), and MNiSW grant DIR/WK/2018/12. Z.S. is supported by the
Department of Science and Technology, Government of India, under
the INSPIRE Faculty grant (DST/INSPIRE/04/2020/002319).

\section*{Data Availability}

The data and software used in this research are available at Fermi Science Support Center (FSSC), NASA's HEASARC webpages with the links given in the manuscript.



\bibliographystyle{mnras}
\bibliography{reference-list.bib}
\bsp	
\label{lastpage}
\end{document}